\documentclass[a4paper,11pt]{article}

\usepackage{amsmath,amssymb,bbm}
\usepackage{ascmac}
\usepackage{graphicx}
\usepackage[bookmarks=true,bookmarksnumbered=true,bookmarkstype=toc]{hyperref}
\hypersetup{
pdftitle={T-duality transformation of GLSM with F-term},
pdfauthor={Tetsuji Kimura and Masaya Yata},
colorlinks={true},
linkcolor={blue},
urlcolor={blue},
filecolor={blue},
citecolor={blue}
}
\makeatletter

 \@addtoreset{equation}{section}
\makeatother

\newcounter{Enumerate}

\DeclareFontFamily{U}{rsf}{}
\DeclareFontShape{U}{rsf}{m}{n}{
  <5> <6> rsfs5 <7> <8> <9> rsfs7 <10-> rsfs10}{}
\DeclareMathAlphabet\Scr{U}{rsf}{m}{n}

\usepackage[mathscr]{eucal}

\newcommand{\del}{\partial}
\newcommand{\half}{\frac{1}{2}}

\newcommand{\mfk}{\mathfrak}
\newcommand{\LS}{\ \ \ \ \ \ \ \ \ \ }
\newcommand{\ls}{\ \ \ \ \ }
\newcommand{\wt}{\widetilde}

\newcommand{\ve}{\varepsilon}
\newcommand{\ol}{\overline}

\newcommand{\bsubeq}{\begin{subequations}}
\newcommand{\esubeq}{\end{subequations}}

\newcommand{\noi}{\noindent}

\newcommand{\nn}{\nonumber}

\newcommand{\A}{\mathscr{A}}

\newcommand{\F}{\mathcal{F}}

\renewcommand{\H}{\mathscr{H}}
\newcommand{\K}{\mathscr{K}}

\newcommand{\N}{\mathcal{N}}

\newcommand\T{\mathcal{T}}
\newcommand{\U}{{\tt U}}
\newcommand{\V}{{\tt V}}
\newcommand{\X}{{\tt X}}
\newcommand{\Y}{{\tt Y}}

\newcommand{\Z}{{\tt Z}}

\renewcommand{\a}{\mfk{a}}
\renewcommand{\d}{{\rm d}}
\newcommand{\e}{{\rm e}}

\renewcommand{\i}{{\rm i}}

\newcommand{\slb}{\scalebox}

\def\+{{+\!\!\!+}} 

\begin{document}
\allowdisplaybreaks{

\thispagestyle{empty}


\begin{flushright}
KEK-TH-1739,
TIT/HEP-637 
\end{flushright}

\vspace{35mm}

\noi
\slb{2.3}{T-duality Transformation of}

\vspace{5mm}

\noi
\slb{2.3}{Gauged Linear Sigma Model with F-term}

\vspace{15mm}

\slb{1.2}{Tetsuji {\sc Kimura}$^{\,a}$} \ and \ \slb{1.2}{Masaya {\sc Yata}$^{\,b}$}

\vspace{3mm}

\slb{.85}{\renewcommand{\arraystretch}{1.2}
\begin{tabular}{rl}
$a$ & {\sl
Department of Physics,
Tokyo Institute of Technology} 
\\
& {\sl Tokyo 152-8551, JAPAN}
\\
& {\tt tetsuji \_at\_ th.phys.titech.ac.jp}
\end{tabular}
}

\slb{.85}{\renewcommand{\arraystretch}{1.2}
\begin{tabular}{rl}
$b$ & {\sl Theory Center, Institute of Particle and Nuclear Studies, KEK}
\\
& {\sl Tsukuba, Ibaraki 305-0801, JAPAN}
\\
& {\tt yata \_at\_ post.kek.jp}
\end{tabular}
}

\vspace{20mm}


\noindent
\slb{1.1}{\sc Abstract}:
\begin{center}
\slb{.95}{
\begin{minipage}{.9\textwidth}
We develop the duality transformation rules in two-dimensional theories in the superfield formalism.
Even if the chiral superfield which we dualize 
involves an F-term, 
we can dualize it
by virtue of the property of chiral superfields.
We apply the duality transformation rule of the neutral chiral superfield to the ${\cal N}=(4,4)$ gauged linear sigma model for five-branes.
We also investigate the duality transformation rule of the charged chiral superfield in the ${\cal N} = (4,4)$ gauged linear sigma model for the $A_1$-type ALE space.
In both cases we obtain the dual Lagrangians in the superfield formalism.
In the low energy limit we find that their duality transformations are interpreted as 
T-duality transformations consistent with the Buscher rule.
\end{minipage}
}
\end{center}

\newpage
\section{Introduction}
\label{sect:introduction}

String worldsheet theory is described in the framework of nonlinear sigma models (NLSMs) or conformal field theories (CFTs)
in the IR regime. 
A gauged linear sigma model (GLSM) \cite{Witten:1993yc} is regarded as the UV completion of a corresponding string worldsheet theory.
Compared with NLSMs and CFTs, GLSMs involve rich structures supported by gauge symmetries.
There are various phases and various branches which characterize the moduli space of the system. 
One can read physically non-trivial phenomena at each point of the moduli space. 
One often finds non-trivial relations between two different points in the moduli space which predict a new feature in string theory.
The reproduction of the Calabi-Yau/Landau-Ginzburg correspondence \cite{Vafa:1988uu} is one of the most striking feature of GLSMs.
One can also discuss duality relations among different GLSMs.
Typical relations are the T-duality transformation \cite{Buscher:1987, Rocek:1991ps}
and mirror symmetry \cite{Strominger:1996it, Hori:2000kt, Mirror03}.

The constituents of the GLSM are 
D-terms, F-terms, and twisted F-terms.
In the conventional GLSM, 
the twisted F-term is often described by 
a twisted chiral superfield converted from a real vector superfield.
This term is significant for analyzing
quantum instanton corrections to the system 
because it yields a topological term given by the gauge field strengths.
The D-term governs the kinematics of 
fields and plays a central role in duality transformations. 
This is because the D-term intrinsically gives
derivative terms, which preserve shift symmetries of fields automatically.
On the other hand, 
the F-term provides non-derivative terms as interactions among the fields.
Such non-derivative terms obstruct T-duality transformations.
However, the existence of an F-term is required
to analyze the NLSM in the IR limit.
The D-term and the F-term in the GLSM respectively
govern the K\"{a}hler structure and 
the complex structure of the target space geometry of the NLSM in the IR limit.
If one considers a string sigma model and its T-duality transformation on a non-trivially curved geometry,
the F-term should be involved in the formulation and in the duality transformation procedure.
Indeed, we advocated the importance of 
F-terms
and constructed the GLSM with F-terms for A-type ALE space \cite{Kimura:2014bxa}. 
In this work, we continue to pursue the importance of F-terms
in GLSMs and develop the consistent duality transformation rules.

\vspace{3mm}

The structure of the paper is as follows.
In section \ref{sect:DTR} we discuss various duality transformation rules of chiral superfields.
If neutral or charged chiral superfields only appear in D-terms,
one can dualize them to twisted chiral superfields in terms of the formula by Ro\v{c}ek and Verlinde \cite{Rocek:1991ps}, which was analyzed by Hori and Vafa \cite{Hori:2000kt}.
We further develop the duality transformation rule of neutral chiral superfields for both the D-term and {\it F-term}. 
This has been exploited in \cite{Kimura:2013fda}.
Next, we propose a new duality transformation rule of {\it charged} chiral superfields involved in both the D-term and {\it F-term}.
In section \ref{sect:EB} 
we discuss a specific example of $\N=(4,4)$ supersymmetric gauge theory, 
which is the GLSM for five-branes.
Here we apply the techniques in section \ref{sect:DTR}.
This has also been discussed in \cite{Tong:2002rq} and \cite{Kimura:2013fda}.
In section \ref{sect:TA-ALE}
we investigate the GLSM for $A_1$-type ALE space \cite{Eguchi:1978gw} and its T-duality.
We follow the toric data of the $A$-type ALE space discussed in \cite{Mirror03}. 
Then we apply the new technique discussed in \ref{subsect:DT-CCS-DF}.
Section \ref{sect:summary} contains the summary and discussions.
In appendix \ref{app:conventions} we write down
the conventions on two-dimensional $\N=(2,2)$ superfield formalism.
They are the basic constituents to formulate two-dimensional $\N=(4,4)$ supersymmetric gauge theories.
In appendix \ref{app:expand} we discuss the explicit form of the GLSM in terms of the bosonic component fields.
We utilize this in section \ref{sect:GLSMF-TALE}.
In appendix \ref{app:EH} we discuss the configurations of the $A_1$-type ALE space and its T-duality.
We also mention the coordinate transformations to the configuration in which two parallel five-branes can be easily recognized.

\section{Duality transformations}
\label{sect:DTR}

In this section we study four types of duality transformations as general discussions: 
the duality transformation for a neutral chiral superfield 
in the D-term (section \ref{subsect:DT-NCS-D}),
for a charged chiral superfield in the D-term (section \ref{subsect:DT-CCS-D}),
for a neutral chiral superfield in the D-term and F-term (section \ref{subsect:DT-NCS-DF}), and
for a charged chiral superfield in the D-term and F-term (section \ref{subsect:DT-CCS-DF}).
The first two have been utilized in various situations so far, 
the third one is the technique which appears in \cite{Kimura:2013fda} for the first time, 
and the fourth is the new duality transformation which we would like to 
explore in this work.
We also insist that they are important when investigating
the duality transformations of chiral superfields with general interactions.

In this section we discuss the above four types in the two-dimensional $\N=(2,2)$ superfield formalism (for the conventions, see appendix \ref{app:N=22SUSY}).
In order to apply them to $\N=(4,4)$ supersymmetric system,
we further impose $SU(2)_R$ symmetry on the supermultiplets.

\subsection{Duality transformation of neutral chiral superfield in D-term}
\label{subsect:DT-NCS-D}

We begin with the simplest Lagrangian of a single free neutral chiral superfield $\Psi$,
\begin{align}
\Scr{L} \ &= \ 
\int \d^4 \theta \, |\Psi|^2
\, . \label{LD-NCS}
\end{align}
Expanding this in terms of the component fields, 
we obtain a free sigma model of a complex scalar field and a Dirac fermion with derivatives.
This belongs to the second order formalism.
Introducing a real superfield $R$ and a twisted chiral superfield $Y$ as auxiliary fields,
we convert the Lagrangian (\ref{LD-NCS}) 
to the first order formalism,
\begin{align}
\Scr{L} \ &= \ 
\int \d^4 \theta \, \Big\{
\half R^2 - \beta R (Y + \ol{Y})
\Big\}
\, . \label{LD-NCS2}
\end{align}
This Lagrangian contains two different models: 
one is the Lagrangian of a dynamical chiral superfield $\Psi$, 
while the other is another Lagrangian of a dynamical twisted chiral superfield $Y$ with a constant $\beta$.
We briefly demonstrate their derivations.
First, we consider the equation of motion for the auxiliary twisted chiral superfield $Y$ in (\ref{LD-NCS2}).
The equation provides a constraint on $R$ 
of $0 = \ol{D}{}_+ D_- R = D_+ \ol{D}{}_- R$. 
We can solve the constraint by using a chiral superfield $\wt{\Psi}$ and its hermitian conjugate in such a way that
$R = \wt{\Psi} + \ol{\wt{\Psi}}$.
Substituting this into (\ref{LD-NCS2}) and identifying $\wt{\Psi}$ with $\Psi$,
we obtain the original Lagrangian (\ref{LD-NCS}).

Next, we go back to the Lagrangian (\ref{LD-NCS2}). 
Instead of considering the field equation for $Y$, 
we study the equation of motion for $R$.
The solution is $R = \beta (Y + \ol{Y})$.
Plugging this into (\ref{LD-NCS2}), 
we obtain the Lagrangian for the twisted chiral superfield $Y$,
\begin{align}
\Scr{L} \ &= \ 
- \int \d^4 \theta \, \beta^2 |Y|^2
\, . \label{DT-LD-NCS}
\end{align}
Here $Y$ is a dynamical superfield because the component fields of $Y$ in this Lagrangian have kinetic terms.
The constant $\beta$ is chosen such that the kinetic terms are in canonical form.
Now we recognize that the Lagrangian (\ref{LD-NCS2}) in the first order formalism relates the two distinct models (\ref{LD-NCS}) and (\ref{DT-LD-NCS}) under the following duality relation,
\begin{align}
\Psi + \ol{\Psi} 
\ &= \ 
\beta (Y + \ol{Y})
\, . \label{PsiY-NCS}
\end{align}
The left-hand side is given by the chiral superfield $\Psi$, 
while the right-hand side is given by the twisted chiral superfield $Y$.
Due to the twist of derivative terms in $Y$ compared with those in $\Psi$,
this relation shows an abelian T-duality transformation 
on the target spaces of (\ref{LD-NCS}) to 
that of (\ref{DT-LD-NCS}) described by the Buscher rule \cite{Buscher:1987}, and vice versa.
This duality transformation appears in a huge amount of works.
In this work we focus only on \cite{Hori:2000kt, Tong:2002rq, Harvey:2005ab, Okuyama:2005gx} exhibited in section \ref{sect:EB} and section \ref{sect:TA-ALE}.

\subsection{Duality transformation of charged chiral superfield in D-term}
\label{subsect:DT-CCS-D}

Next, we introduce an abelian gauge symmetry into the original system (\ref{LD-NCS}),
\begin{align}
\Scr{L} \ &= \ 
\int \d^4 \theta \, |\Psi|^2 \, \e^{+ 2 \alpha V}
\, . \label{LD-CCS}
\end{align}
Here we focus on a chiral superfield $\Psi$ of charge $+ \alpha$. 
We neglect the kinetic term of the abelian vector superfield $V$ 
because it does not contribute to the duality transformation.
The Lagrangian is 
in the second order formalism because the bosonic component fields are governed by second derivatives in a gauge covariant way.
We convert (\ref{LD-CCS}) to 
a first order Lagrangian,
\begin{align}
\Scr{L} \ &= \ 
\int \d^4 \theta \, \Big\{
\e^{2 \alpha V + R}
- \beta R (Y + \ol{Y})
\Big\}
\, , \label{LD-CCS2}
\end{align}
where $R$ and $Y$ are an auxiliary real superfield and an auxiliary twisted chiral 
superfield respectively.
$\beta$ is a constant determined by the normalization of the kinetic term in the dual theory.
Due to the gauge invariance, the contribution of $R$ to the Lagrangian is slightly different from the one in (\ref{LD-NCS2}).
If we consider the equation of motion for $Y$, 
we obtain the constraint $0 = \ol{D}{}_+ D_- R = D_+ \ol{D}{}_- R$, 
whose solution is again given by a chiral superfield $\wt{\Psi}$ as $R = \wt{\Psi} + \ol{\wt{\Psi}}$.
Substituting this into (\ref{LD-CCS2}) and identifying $\e^{\wt{\Psi} + \ol{\wt{\Psi}}}$ with $|\Psi|^2$, we obtain the original Lagrangian (\ref{LD-CCS}).
If we consider the equation of motion for $R$ instead of the equation of motion for $Y$, 
we obtain $0 = \e^{2 \alpha V + R} - \beta (Y + \ol{Y})$.
Plugging this into (\ref{LD-CCS2}), 
we obtain the dual Lagrangian of the twisted chiral superfield $Y$,
\begin{align}
\Scr{L} \ &= \ 
\int \d^4 \theta \, \Big\{
- \beta (Y + \ol{Y}) \Big( -2 \alpha V + \log \beta (Y + \ol{Y}) 
\Big)
\Big\}
\nn \\
\ &= \ 
\int \d^4 \theta \, \Big\{
- \beta (Y + \ol{Y}) \log (Y + \ol{Y})
\Big\}
+ \Big\{ \sqrt{2} \int \d^2 \wt{\theta} \, \alpha \beta \, Y \Sigma
+ \text{(h.c.)}
\Big\}
\, . \label{DT-LD-CCS}
\end{align}
Here we introduced another twisted chiral superfield $\Sigma$ given by the vector superfield such that
$\Sigma = \frac{1}{\sqrt{2}} \ol{D}{}_+ D_- V$.
The second term of the right-hand side is a twisted F-term, which is converted from the D-term in the following way,
\begin{align}
\int \d^4 \theta \, \beta (Y + \ol{Y}) (2 \alpha V)
\ &= \ 
\Big\{
\sqrt{2} \int \d^2 \wt{\theta} \, \alpha \beta \, Y \Sigma
+ \text{(h.c.)}
\Big\}
+ \text{(total derivative terms)}
\, . \label{V-Sigma}
\end{align}
If we restrict ourselves to a classical string sigma model without worldsheet boundaries, we can ignore the total derivative terms.
Expanding (\ref{DT-LD-CCS}) in terms of the component fields of $Y$, 
we find that the Lagrangian (\ref{DT-LD-CCS}) contains the kinetic terms of them.
The relation between $\Psi$ in (\ref{LD-CCS}) and $Y$ in (\ref{DT-LD-CCS}) is determined via the auxiliary superfield $R$ as
\begin{align}
|\Psi|^2 \, \e^{+2 \alpha V} 
\ &= \ 
\beta (Y + \ol{Y})
\, . \label{PsiY-CCS}
\end{align}
This is a generalization of the T-duality transformation (\ref{PsiY-NCS}) including an abelian gauge symmetry.
We also understand that the Lagrangian (\ref{LD-CCS2}) in the first order formalism is a generalization of (\ref{LD-NCS2}).
This duality transformation appears in \cite{Rocek:1991ps, Hori:2000kt}.
Indeed this duality transformation rule is one of the most useful ones in two-dimensional $\N=(2,2)$ supersymmetric theories.
Various applications and discussions can be seen in \cite{Mirror03}.

So far, 
we demonstrated two famous duality transformation rules for chiral superfields involved only in the D-term.
In the next two subsections, 
we investigate the duality transformation rules for interacting chiral superfields in D-terms and F-terms.

\subsection{Duality transformation of neutral chiral superfield in D-term and F-term}
\label{subsect:DT-NCS-DF}

We generalize the Lagrangian of the single free neutral chiral superfield (\ref{LD-NCS}) by introducing an F-term 
given by a superpotential $- \Phi {\cal W} \, \Psi$, 
where $\Phi$ is an arbitrary neutral chiral superfield 
and ${\cal W}$ is an arbitrary holomorphic function of arbitrary chiral superfields.
The Lagrangian is 
\begin{align}
\Scr{L} \ &= \ 
\int \d^4 \theta \, |\Psi|^2
+ \Big\{ \sqrt{2} \int \d^2 \theta \, (- \Phi) {\cal W} \, \Psi
+ \text{(h.c.)}
\Big\} 
\, . \label{LDF-NCS}
\end{align}
It seems that the form of the superpotential $- \Phi {\cal W} \, \Psi$ is special.
However, such a form often appears in the consideration of the GLSMs for Calabi-Yau varieties as hypersurfaces of toric geometries (see, for instance, \cite{Witten:1993yc, Mirror03}), 
and in $\N=(4,4)$ supersymmetric systems (see, for instance, \cite{Tong:2002rq, Kimura:2013fda}).

We convert the F-term to D-terms using a property of chiral superfields.
Since the chiral superfield $\Phi$ is described in terms of an unconstrained complex superfield $C$ in such a way that $\Phi = \ol{D}{}_+ \ol{D}{}_- C$, 
the F-term in (\ref{LDF-NCS}) can be converted to the following D-terms,
\begin{align}
&\sqrt{2} \int \d^2 \theta \, (- \Phi) {\cal W} \, \Psi
+ \text{(h.c.)}
\nn \\
\ &\LS \ \ = \ 
- \sqrt{2} \int \d^4 \theta \, \Big\{
(\Psi + \ol{\Psi}) ({\cal W} C + \ol{\cal W} \ol{C}) + (\Psi - \ol{\Psi}) ({\cal W} C - \ol{\cal W} \ol{C})
\Big\}
\, . \label{F2D-NCS}
\end{align}
Introducing two auxiliary real superfields $\{ R, S \}$, two auxiliary twisted chiral superfields $\{ Y, Y' \}$, and an auxiliary chiral superfield $X$,
we lift up (\ref{LDF-NCS}) to the following Lagrangian in the first order 
formalism,
\begin{align}
\Scr{L} \ &= \ 
\int \d^4 \theta \, \Big\{
\frac{1}{2} R^2
+ \beta R (Y + \ol{Y})
- \sqrt{2} \, R ({\cal W} C + \ol{\cal W} \ol{C})
+ R (X + \ol{X})
\Big\}
\nn \\
\ & \ \ \ \ 
+ \int \d^4 \theta \, \Big\{
\beta (\i S) (Y' - \ol{Y}{}')
- \sqrt{2} \, (\i S) ({\cal W} C - \ol{\cal W} \ol{C})
+ (\i S) (X - \ol{X})
\Big\}
\, , \label{LDF-NCS2}
\end{align}
where $\beta$ is an arbitrary constant.
Since the imaginary part of ${\cal W} C$ is involved in the D-terms (\ref{F2D-NCS}), 
we introduced additional auxiliary fields $\{ S, Y', X \}$ compared with the system (\ref{LD-NCS2}).

We should investigate this Lagrangian and the duality transformation rule by integrating out the auxiliary superfields.
First, we consider the reduction of (\ref{LDF-NCS2}) to the original Lagrangian (\ref{LDF-NCS}).
Integrating out the auxiliary superfields $\{ Y, Y' \}$, 
we obtain the constraints on $\{ R , S \}$ and the solution, such as
\bsubeq \label{const-RS-NCS}
\begin{alignat}{3}
\ol{D}{}_+ D_- R 
\ &= \ 
0 \ = \ D_+ \ol{D}{}_- R
\ \ \ &&\to &\ \ \ 
R \ &= \ 
\wt{\Psi}_1 + \ol{\wt{\Psi}}{}_1
\, , \\
\ol{D}{}_+ D_- (\i S) 
\ &= \ 
0 \ = \ D_+ \ol{D}{}_- (\i S)
\ \ \ &&\to &\ \ \ 
\i S \ &= \ \wt{\Psi}_2 - \ol{\wt{\Psi}}{}_2
\, . 
\end{alignat}
\esubeq
Further integrating out $X$ under the constraints (\ref{const-RS-NCS}), 
we find that the two chiral superfields $\wt{\Psi}_1$ and $\wt{\Psi}_2$ are identical with each other.
Substituting the result $R = \wt{\Psi}_1 + \ol{\wt{\Psi}}_1 
= \wt{\Psi}_2 + \ol{\wt{\Psi}}_2$ 
with the identification $\wt{\Psi}_{1} = \wt{\Psi}_2 \equiv \Psi$ into the first order Lagrangian (\ref{LD-NCS2}),
we obtain the original Lagrangian (\ref{LDF-NCS}).

Second, we construct the dual Lagrangian from (\ref{LDF-NCS2}).
Integrating out $R$ and $Y'$, we obtain two constraints,
\bsubeq
\begin{align}
0 \ &= \ 
R + \beta (Y + \ol{Y}) - \sqrt{2} \ ({\cal W} C + \ol{\cal W} \ol{C})
+ (X + \ol{X})
\nn \\
\ &= \ 
R + \beta (Y + \ol{Y}) - \sqrt{2} \ ({\cal W} C' + \ol{\cal W} \ol{C}{}')
\, , \\
0 \ &= \ 
\ol{D}{}_+ D_- (\i S) \ = \ 
D_+ \ol{D}{}_- (\i S)
\ \ \ \to \ \ \ 
\i S \ = \ 
\wt{\Psi}_2 - \ol{\wt{\Psi}}{}_2
\, .
\end{align}
\esubeq
Here we absorbed the auxiliary chiral superfield $X$ into the unconstrained complex superfield $C$ without loss of generality.
Plugging this into (\ref{LDF-NCS2}), 
we obtain 
\begin{align}
\Scr{L} \ &= \ 
\int \d^4 \theta \, \Big\{
- \frac{1}{2} \Big( \beta (Y + \ol{Y}) - \sqrt{2} \, ({\cal W} C + \ol{\cal W} \ol{C}) \Big)^2 
- \sqrt{2} \, (\Psi - \ol{\Psi}) ({\cal W} C - \ol{\cal W} \ol{C})
\Big\}
\, . \label{DT-LDF-NCS}
\end{align}
Note that we neglected the prime attached to the superfield $C$ 
because all the degrees of freedom 
of the auxiliary chiral superfield $X$ are absorbed into those of $C$. 
This is the dual Lagrangian of the original one (\ref{LDF-NCS}).
The duality relation between the chiral superfield $\Psi$ in (\ref{LDF-NCS}) and the twisted chiral superfield $Y$ in (\ref{DT-LDF-NCS}) is given via the equations of motion for $\{ R , S \}$,
\bsubeq \label{PsiXi-NCS}
\begin{align}
\Psi + \ol{\Psi}
\ &= \ 
- \beta (Y + \ol{Y})
+ \sqrt{2} \, ({\cal W} C + \ol{\cal W} \ol{C})
\, , \\
\Psi - \ol{\Psi}
\ &= \
\wt{\Psi}_2 - \ol{\wt{\Psi}}{}_2
\, . 
\end{align}
\esubeq
The most important ingredient of the dual Lagrangian is the last term in (\ref{DT-LDF-NCS}).
This involves the imaginary part of the original chiral superfield $\Psi$, which is no longer dynamical.
The existence of this term is the most crucial point to complete the T-duality transformation on the target space configuration.
We should notice that this term generally breaks a global shift symmetry of the imaginary part of the scalar field of $\Psi$.
This implies that there is no 
isometry.
Thus, strictly speaking, we cannot perform a T-duality transformation.
In order to avoid this difficulty, we have to use a trick.
In section \ref{subsect:522}, we will demonstrate this duality transformation rule with a trick in a concrete way, which has been evaluated in \cite{Kimura:2013fda, Kimura:2013zva, Kimura:2013khz}.

We have two comments on this system.
The first comment is the possibility of other combinations of auxiliary superfields which could be integrated out.
However, all other possibilities yield transformations that are
either intrinsically the same as the above two,
or are inconsistent with the duality transformations 
on the target space configuration.
Another comment is on the Lagrangian containing twisted F-terms.
If the twisted F-term is of the form $Y \Sigma$ with $\Sigma = \frac{1}{\sqrt{2}} \ol{D}{}_+ D_- V$,
we can always convert it to a D-term as in (\ref{PsiY-CCS}).
In this case the imaginary part of $Y \Sigma$ does not contribute to the system 
because $V$ is a {\it real} vector superfield.
This situation can be seen in section \ref{subsect:KKM}.
However, 
if we consider a Lagrangian of a twisted F-term given by arbitrary twisted chiral superfields, 
the imaginary part of it does contribute to the system.
In this case we again can formulate a first order Lagrangian similar to (\ref{LDF-NCS}).
We do not seriously consider this case in this work.

\subsection{Duality transformation of charged chiral superfield in D-term and F-term}
\label{subsect:DT-CCS-DF}

Finally we consider the Lagrangian of a single charged chiral superfield with a superpotential $- \wt{\alpha} \Phi {\cal W} \, \Psi$ in the F-term.
$\Phi$ is again an arbitrary neutral chiral superfield,
${\cal W}$ is an arbitrary holomorphic function of arbitrary chiral superfields,
and $\Psi$ is a chiral superfield of charge $+ \alpha$ under an abelian gauge symmetry given by the vector superfield $V$.
The Lagrangian is
\begin{align}
\Scr{L} \ &= \ 
\int \d^4 \theta \, 
\Big( |\Psi|^2 \, \e^{+2 \alpha V} \Big)
+ \Big\{ \sqrt{2} \int \d^2 \theta \, (- \wt{\alpha} \Phi) {\cal W} \, \Psi
+ \text{(h.c.)}
\Big\}
\, , \label{LDF-CCS}
\end{align}
where $\wt{\alpha}$ is an arbitrary constant in a generic $\N=(2,2)$ supersymmetric theory.
This is a generalization of the Lagrangian of the single charged chiral superfield (\ref{LD-CCS}), or 
a generalization of the Lagrangian of the single neutral chiral superfield with the F-term interactions (\ref{LDF-NCS}).
If we study $\N=(4,4)$ theories, 
we have to impose $\wt{\alpha} = \alpha$ in order to preserve $SU(2)_R$ symmetry.

Since the neutral chiral superfield $\Phi$ is described by an unconstrained complex superfield $\Phi = \ol{D}{}_+ \ol{D}{}_- C$, 
we convert the F-term in (\ref{LDF-CCS}) to D-terms in the following way,
\begin{align}
- \sqrt{2} \int \d^2 \theta \, \wt{\alpha} \Phi {\cal W} \, \Psi
+ \text{(h.c.)}
\ &= \ 
- 2 \sqrt{2} \, \wt{\alpha} \int \d^4 \theta \, \Big\{
\Psi {\cal W} C + \ol{\Psi} \ol{\cal W} \ol{C}
\Big\}
\, .
\end{align}
Thus we can construct the following Lagrangian in the first order formalism,
\begin{align}
\Scr{L} \ &= \ 
\int \d^4 \theta \, \Big\{
\e^{+2 \alpha V + R}
- \beta R (Y + \ol{Y})
- 2 \sqrt{2} \, \wt{\alpha} \Big[
\e^{\frac{1}{2} (R + \i S)} {\cal W} C
+ \e^{\frac{1}{2} (R - \i S)} \ol{\cal W} \ol{C}
\Big] 
\Big\}
\nn \\
\ & \ \ \ \ 
+ \int \d^4 \theta \, \Big\{
- \beta (\i S) (Y' - \ol{Y}{}')
- \e^{\half (R + \i S)} X 
- \e^{\half (R - \i S)} \ol{X} 
\Big\}
\, . \label{LDF-CCS2}
\end{align}
Following the same discussion of (\ref{LDF-NCS2}), we introduced 
two auxiliary real superfields $\{ R , S \}$,
two auxiliary twisted chiral superfields $\{ Y, Y' \}$,  
an auxiliary chiral superfield $X$,
and an arbitrary constant $\beta$.

First, we derive the original Lagrangian (\ref{LDF-CCS}).
Integrating out the auxiliary superfields $\{ Y, Y' \}$, 
we obtain the constraints and the solution of $\{ R, S \}$,
\bsubeq
\begin{alignat}{3}
\ol{D}{}_+ D_- R \ &= \ 0 \ = \ 
D_+ \ol{D}{}_- R
\ \ \ &\to &&\ \ \ 
R \ &= \ \wt{\Psi}_1 + \ol{\wt{\Psi}}{}_1
\, , \\
\ol{D}{}_+ D_- (\i S) \ &= \ 0 \ = \ 
D_+ \ol{D}{}_- (\i S)
\ \ \ &\to &&\ \ \ 
\i S \ &= \ \wt{\Psi}_2 - \ol{\wt{\Psi}}{}_2
\, .
\end{alignat}
\esubeq
Under the constraints we further integrate out $X$,
finding another set of constraints,
\bsubeq
\begin{align}
0 \ = \ 
\ol{D}{}_+ \ol{D}{}_- (\e^{\half (R + \i S)})
\ &= \ 
\ol{D}{}_+ \ol{D}{}_- \exp \Big[
\half (\wt{\Psi}_1 + \wt{\Psi}_2)
+ \half (\ol{\wt{\Psi}}{}_1 - \ol{\wt{\Psi}}{}_2)
\Big]
\nn \\
\ &= \
\ol{D}{}_+ \ol{D}{}_- \exp \Big[
\half (\ol{\wt{\Psi}}{}_1 - \ol{\wt{\Psi}}{}_2)
\Big] 
\, , \\
0 \ = \ 
D_+ D_- (\e^{\half (R - \i S)})
\ &= \ 
D_+ D_- \exp \Big[
\half (\wt{\Psi}_1 - \wt{\Psi}_2)
+ \half (\ol{\wt{\Psi}}{}_1 + \ol{\wt{\Psi}}{}_2)
\Big]
\nn \\
\ &= \ 
D_+ D_- \exp \Big[
\half (\wt{\Psi}_1 - \wt{\Psi}_2)
\Big]
\, .
\end{align}
\esubeq
It turns out that the two chiral superfields $\wt{\Psi}_1$ and $\wt{\Psi}_2$ are identical with each other.
Plugging the solution into (\ref{LDF-CCS2}) with the identification $\e^{\wt{\Psi}_1} = \e^{\wt{\Psi}_2} \equiv \Psi$, we find the original Lagrangian (\ref{LDF-CCS}).

Second, we derive the dual Lagrangian from the first order Lagrangian (\ref{LDF-CCS2}).
Integrating out $R$ and $Y'$, we obtain the following two constraints,
\bsubeq
\begin{align}
0 \ &= \ 
\e^{2 \alpha V + R}
- \beta (Y + \ol{Y})
- \sqrt{2} \, \wt{\alpha} \Big[
\e^{\half (R + \i S)} {\cal W} C
+ \e^{\half (R - \i S)} \ol{\cal W} \ol{C}
\Big]
\nn \\
\ & \ \ \ \ 
- \half \e^{\half (R + \i S)} X
- \half \e^{\half (R - \i S)} \ol{X}
\nn \\
\ &= \ 
\e^{2 \alpha V + R}
- \beta (Y + \ol{Y})
- \sqrt{2} \, \wt{\alpha} \Big[
\e^{\half (R + \i S)} {\cal W} C'
+ \e^{\half (R - \i S)} \ol{\cal W} \ol{C}{}'
\Big]
\, , \\
0 \ &= \ 
\ol{D}{}_+ D_- (\i S) \ = \ 
D_+ \ol{D}{}_- (\i S)
\, .
\end{align}
\esubeq
Here $X$ is absorbed into $C$ without loss of generality.
The solution of $\{ R, S \}$ is given as
\bsubeq
\begin{align}
R \ &= \ 
- 2 V + 2 \log \Big\{
- \T + \sqrt{\T^2 + 4 \beta (Y + \ol{Y})}
\Big\}
- 2 \log 2
\, , \\
\T \ &\equiv \ 
- \sqrt{2} \, \wt{\alpha} \, \e^{- V} \, \Big\{
\e^{\frac{\i}{2} S} \, {\cal W} C'
+ \e^{- \frac{\i}{2} S} \, \ol{\cal W} \ol{C}{}'
\Big\}
\, , \\
\i S \ &= \ 
\wt{\Psi} - \ol{\wt{\Psi}}
\, .
\end{align}
\esubeq
Plugging this into (\ref{LDF-CCS2}), we obtain the dual Lagrangian,
\bsubeq \label{DT-LDF-CCS}
\begin{align}
\Scr{L} \ &= \ 
\int \d^4 \theta \, \Big\{
- 2 \beta (Y + \ol{Y}) \log \F
+ \half \F \T
\Big\}
+ \Big\{
\sqrt{2} \int \d^2 \wt{\theta} \, \beta Y \Sigma
+ \text{(h.c.)}
\Big\}
\, , \label{DT-LDF-CCS-1} \\
\F \ &= \ 
- \T + \sqrt{\T^2 + 4 \beta (Y + \ol{Y})}
\, , \\
\T \ &= \ 
- \sqrt{2} \, \wt{\alpha} \, \e^{- V} \, \Big\{
\e^{\half (\wt{\Psi} - \ol{\wt{\Psi}})} \, {\cal W} C
+ \e^{- \half (\wt{\Psi} - \ol{\wt{\Psi}})} \, \ol{\cal W} \ol{C}
\Big\}
\, , \\
\Phi \ &= \ 
\ol{D}{}_+ \ol{D}{}_- C 
\, . 
\end{align}
\esubeq
Here we again removed the prime attached with the superfield $C'$ 
because the degrees of freedom
of the chiral superfield $X$ can be absorbed into $C$ completely.
We also used the conversion (\ref{V-Sigma}).
Finally we show the duality relation 
between the chiral superfield $\Psi$ in (\ref{LDF-CCS}) 
and the twisted chiral superfield $Y$ in (\ref{DT-LDF-CCS}) 
in the presence of the unconstrained complex superfield $C$,
\bsubeq \label{AY-dual-CCS}
\begin{align}
|\Psi|^2 \, \e^{+ 2 \alpha V} 
\ &= \ 
\frac{\F^2}{4}
\ = \ 
\beta (Y + \ol{Y})
+ \frac{\T^2}{2} 
- \frac{\T}{2} \sqrt{\T^2 + 4 \beta (Y + \ol{Y})}
\, , \\
\Psi \ &= \ \e^{\wt{\Psi}}
\, . 
\end{align}
\esubeq
Note that we prepare 
explicit expressions of the first term in (\ref{DT-LDF-CCS-1}) and of (\ref{AY-dual-CCS}) in terms of the component fields in appendix \ref{app:expand}.
There are a huge number of bosonic terms, 
even though we neglect all the fermionic terms.
The explicit expressions would be necessary in order to investigate the whole structure of the moduli space of the dual Lagrangian (\ref{DT-LDF-CCS}).

We have another comment on the novel duality relation (\ref{AY-dual-CCS}).
If the superfield $C$ vanishes, the dual Lagrangian (\ref{DT-LDF-CCS}) is reduced to the dual Lagrangian (\ref{DT-LD-CCS}).
This is similar to how the original Lagrangian (\ref{LDF-CCS}) is reduced to (\ref{DT-LD-CCS}).
On the other hand, we should keep in mind that some component fields of the superfield $C$ can be still non-trivial, even though the chiral superfield $\Phi$ vanishes.

\section{GLSM for five-branes}
\label{sect:EB}

In ten-dimensional string theory
there exist various kinds of five-branes:
an NS5-brane (or an H-monopole as a smeared NS5-brane along one direction), a Kaluza-Klein (KK) monopole,
and an exotic $5^2_2$-brane. 
These are three typical examples\footnote{Other standard or exotic five-branes in string theories and M-theory are discussed in section 2 of \cite{Kimura:2014upa}.}.
They are closely related to each other under the T-duality transformations.
In this section we determine the GLSMs for them.
The GLSMs for the NS5-branes and the KK-monopoles are discussed in \cite{Tong:2002rq, Harvey:2005ab, Okuyama:2005gx}, 
while the GLSM for an exotic $5^2_2$-brane is in \cite{Kimura:2013fda, Kimura:2013zva, Kimura:2013khz}.

\subsection{NS5-branes or H-monopoles}
\label{subsect:NS5}

We begin with the GLSM for $k$-centered H-monopoles discussed in \cite{Tong:2002rq, Okuyama:2005gx}. 
The $\N=(4,4)$ Lagrangian itself is in rather a simple form,
\begin{align}
\Scr{L}_{\text{H}} \ &= \ 
\sum_{a=1}^k \int \d^4 \theta \, \Big\{ 
\frac{1}{e_a^2} 
\Big( - |\Sigma_a|^2 + |\Phi_a|^2 \Big)
+ |Q_a|^2 \, \e^{-2 V_a} 
+ |\wt{Q}_a|^2 \, \e^{+2 V_a} 
\Big\}
\nn \\
\ & \ \ \ \
+ \int \d^4 \theta \, \Big\{
\frac{1}{g^2} \Big( |\Psi|^2 - |\Theta|^2 \Big)
\Big\}
\nn \\
\ & \ \ \ \ 
+ \sum_{a=1}^k \Big\{
\sqrt{2} \int \d^2 \theta \, \Big( \wt{Q}_a \Phi_a Q_a + (s_a - \Psi) \Phi_a
\Big)
+ \text{(h.c.)} 
\Big\}
\nn \\
\ & \ \ \ \ 
+ \sum_{a=1}^k \Big\{ 
\sqrt{2} \int \d^2 \wt{\theta} \, (t_a - \Theta) \Sigma_a
+ \text{(h.c.)}
\Big\}
\, . \label{GLSM-HM}
\end{align}
There are various $\N=(4,4)$ supermultiplets\footnote{For the details of conventions, see \cite{Kimura:2013fda}.}:
$k$ vector multiplets $\{ V_a, \Phi_a \}$,
$k$ charged hypermultiplets $\{ Q_a, \wt{Q}_a \}$,
and a neutral hypermultiplet $\{ \Psi, \Theta \}$.
All the multiplets are built with $\N=(2,2)$ superfields 
which are subject to the $SU(2)_R$ symmetry.
Further we introduce the Fayet-Iliopoulos (FI) parameters 
$s_a = \frac{1}{\sqrt{2}} (s_a^1 + \i s_a^2)$ 
and $t_a = \frac{1}{\sqrt{2}} (t_a^3 + \i t_a^4)$.
We also use the definition
$\Sigma_a = \frac{1}{\sqrt{2}} \ol{D}{}_+ D_- V_a$ in the D-term and the twisted F-term.

It is worth expanding the Lagrangian (\ref{GLSM-HM}) in the component fields.
In this procedure we focus only on the bosonic sector.
Integrating out all the auxiliary fields, we find the following form,
\begin{align}
\Scr{L}_{\text{H}}
\ &= \ 
\sum_{a=1}^k \frac{1}{e_a^2} \Big\{
\half (F_{a,01})^2 
- |\del_m \sigma_a|^2
- |\del_m \phi_a|^2
\Big\}
- \frac{1}{2 g^2}
\Big\{ (\del_m \vec{r})^2 + (\del_m \vartheta)^2 \Big\}
\nn \\
\ & \ \ \ \ 
- \sum_{a=1}^k \Big\{
|D_m q_a|^2
+ |D_m \wt{q}_a|^2
\Big\}
- \sqrt{2} \sum_{a=1}^k (\vartheta - t_{a}^4) \, F_{a,01}
\nn \\
\ & \ \ \ \ 
- 2 \sum_{a=1}^k \big( |\sigma_a|^2 + |\phi_a|^2 \big) 
\big( |q_a|^2 + |\wt{q}_a|^2 \big)
- 2 g^2 \sum_{a,b} \big( \sigma_a \ol{\sigma}{}_b + \phi_a \ol{\phi}{}_b \big)
\nn \\
\ & \ \ \ \ 
- \sum_{a=1}^k \frac{e_a^2}{2} 
\Big( |q_a|^2 - |\wt{q}_a|^2 - \sqrt{2} (r^3 - t_{a}^3) \Big)^2
- \sum_{a=1}^k e_a^2 \Big|
\sqrt{2} \, q_a \wt{q}_a - \big( (r^1 - s_{a}^2) + \i (r^2 - s_{a}^2) \big) 
\Big|^2
\nn \\
\ & \ \ \ \ 
+ \text{(fermionic fields)}
\, . \label{GLSM-Hm-b1}
\end{align}
The scalar fields $\{ \sigma_a, \phi_a, q_a, \wt{q}_a \}$
are the scalar component fields of the superfields
$\{ \Sigma_a, \Phi_a, Q_a, \wt{Q}_a \}$ respectively.
The scalar components of $\{ \Psi, \Theta \}$ are given as
$\{ \frac{1}{\sqrt{2}} (r^1 + \i r^2), \frac{1}{\sqrt{2}} (r^3 + \i \vartheta) \}$.
We introduced the gauge covariant derivatives 
$D_m q_a = \del_m q_a - \i A_m q_a$
and $D_m \wt{q}_a = \del_m \wt{q}_a + \i A_m \wt{q}_a$.
We also utilized the expression $\vec{r} = (r^1, r^2, r^3)$.

We discuss the supersymmetric Higgs branch.
This is given by the following constraints,
\bsubeq \label{Higgsbranch-ex1}
\begin{align}
0 \ &= \ 
\sigma_a \ = \ \phi_a
\, , \\
0 \ &= \ 
|q_a|^2 - |\wt{q}_a|^2 - \sqrt{2} \, (r^3 - t_{a}^3) 
\, , \\
0 \ &= \ 
\sqrt{2} \, q_a \wt{q}_a - \big( (r^1 - s_{a}^1) + \i (r^2 - s_{a}^2) \big) 
\, .
\end{align}
\esubeq
The latter two constraints provide a non-trivial solution of $\{ q_a, \wt{q}_a \}$ in terms of $\vec{r}$ and $\{ s_a, t_a \}$.
Substituting the solution into the Lagrangian (\ref{GLSM-Hm-b1}), we find
\begin{align}
\Scr{L}_{\text{H}}
\ &= \ 
\sum_{a=1}^k \frac{1}{2 e_a^2} (F_{a,01})^2
- \frac{1}{2} \Big\{ \frac{1}{g^2} + \sum_{a=1}^k \frac{1}{\sqrt{2} R_a} \Big\} (\del_m \vec{r})^2
- \frac{1}{2 g^2} (\del_m \vartheta)^2 
\nn \\
\ & \ \ \ \ 
- \sum_{a=1}^k \sqrt{2} R_a \Big( 
A_{a,m} + \frac{1}{\sqrt{2}} \, {\omega}_{a,i} \, \del_m r^i 
\Big)^2
- \sqrt{2} \sum_{a=1}^k (\vartheta - t_{a}^4) \, F_{a,01}
\nn \\
\ & \ \ \ \ 
+ \text{(fermionic fields)}
\, , \label{GLSM-Hm-b2}
\end{align}
where we introduced the following functions,
\bsubeq \label{omega-R}
\begin{gather}
\omega_{a,1} \ = \ 
\frac{r^2 - s_{a}^2}{\sqrt{2} R_a (R_a + (r^3 - t_{a}^3))}
\, , \ls
\omega_{a,2} \ = \ 
- \frac{r^1 - s_{a}^1}{\sqrt{2} R_a (R_a + (r^3 - t_{a}^3))}
\, , \ls
\omega_{a,3} \ = \ 0
\, , \\
R_a
\ = \ 
\sqrt{ \vphantom{\big|} (r^1 - s_{a}^1)^2 + (r^2 - s_{a}^2)^2 + (r^3 - t_{a}^3)^2}
\, . 
\end{gather}
\esubeq
In the IR limit, the dimensionful gauge coupling constants $e_a$ 
go to infinity and the kinetic terms of the gauge fields disappear.
Then the gauge fields become auxiliary fields. 
Integrating them out, 
we finally obtain the $\N=(4,4)$ supersymmetric NLSM in the following form,
\bsubeq \label{GLSM-Hm-b3}
\begin{align}
\Scr{L}_{\text{H}}
\ &= \ 
- \half H \Big\{ (\del_m \vec{r})^2 + (\del_m \vartheta)^2 \Big\}
+ \ve^{mn} \, \omega_i \, (\del_m r^i) \, (\del_n \vartheta)
+ \text{(fermionic fields)}
\, , \\
H \ &\equiv \ 
\frac{1}{g^2} + \sum_{a=1}^k \frac{1}{\sqrt{2} R_a}
\, , \ls
\omega_i \ \equiv \ 
\sum_{a=1}^k \omega_{a,i}
\, , \ls
\nabla_i H \ = \ 
(\nabla \times \omega)_i
\, . \label{H-omega-Hm}
\end{align}
\esubeq
We can read off the target space configuration 
by comparing to the string worldsheet sigma model (\ref{string-NLSM-bosons}).
The target space is
expanded by four coordinates associated with the four scalar fields $\{ r^1, r^2, r^3, \vartheta \}$.
The target space metric gives rise to the $k$-centered H-monopoles whose centers are parametrized by 
$k$ sets of the FI parameters $\{ s_a, t_a \}$.
This implies that each set of the $\N=(4,4)$ vector multiplet yields a single five-brane \cite{Okuyama:2005gx}.
The second term in the right-hand side represents the target space B-field.
The function $H$ is the harmonic function which governs the target space metric, the B-field and the dilaton.
Since $H$ does not depend on $\vartheta$, there is an abelian isometry along this direction.
Due to the independence, we refer to the five-brane as the H-monopole, 
which is the smeared NS5-brane.
The target space dilaton does not explicitly appear in the NLSM because the two-dimensional base space in the NLSM is flat.
The functions $H$ and $\omega_i$ satisfy the monopole equation.
This configuration is a solution of ten-dimensional supergravity.

\subsection{KK-monopoles}
\label{subsect:KKM}

As mentioned before, 
the configuration of the KK-monopoles is realized by the T-duality transformation from that of H-monopoles.
This is also realized in the framework of the GLSM \cite{Tong:2002rq, Harvey:2005ab, Okuyama:2005gx}.
In order to dualize the scalar field $\vartheta$ in the NLSM (\ref{GLSM-Hm-b3}) at the GLSM level,
we perform the duality transformation on the twisted chiral superfield $\Theta$ in (\ref{GLSM-HM}).
Following the procedure in section \ref{subsect:DT-NCS-D} with 
a conversion similar to (\ref{V-Sigma}),
we obtain the duality relation on the superfields,
\begin{align}
\frac{1}{g^2} (\Theta + \ol{\Theta})
\ &= \ 
- (\Gamma + \ol{\Gamma}) + 2 \sum_{a=1}^k V_a
\, . \label{Theta-Gamma-KKM}
\end{align}
In the right-hand side the chiral superfield $\Gamma$ is the dualized object.
This is coupled to the vector superfield because of the contribution of the twisted F-term in (\ref{GLSM-HM}).
Subject to this duality relation, we obtain the dual GLSM in the following form,
\begin{align}
\Scr{L}_{\text{KK}} \ &= \ 
\sum_{a=1}^k \int \d^4 \theta \, \Big\{ 
\frac{1}{e_a^2} 
\Big( - |\Sigma_a|^2 + |\Phi_a|^2 \Big)
+ |Q_a|^2 \, \e^{-2 V_a} 
+ |\wt{Q}_a|^2 \, \e^{+2 V_a} 
\Big\}
\nn \\
\ & \ \ \ \
+ \int \d^4 \theta \, \Big\{
\frac{1}{g^2} |\Psi|^2
+ \frac{g^2}{2} 
\Big( \Gamma + \ol{\Gamma} + 2 \sum_{a=1}^k V_a \Big)^2
\Big\}
\nn \\
\ & \ \ \ \ 
+ \sum_{a=1}^k \Big\{
\sqrt{2} \int \d^2 \theta \, \Big( \wt{Q}_a \Phi_a Q_a + (s_a - \Psi) \Phi_a
\Big)
+ \text{(h.c.)} 
\Big\}
\nn \\
\ & \ \ \ \ 
+ \sum_{a=1}^k \Big\{ 
\sqrt{2} \int \d^2 \wt{\theta} \, (t_a \, \Sigma_a)
+ \text{(h.c.)}
\Big\}
- \sqrt{2} \, \ve^{mn} \sum_{a=1}^k \del_m (\vartheta A_{a,n})
\, . \label{GLSM-KKM}
\end{align}
The total derivative term including the gauge fields appears in the conversion of the twisted F-term in (\ref{GLSM-HM}).
In this work we ignore this total derivative term in (\ref{GLSM-KKM}) which plays a central role in quantum corrections \cite{Tong:2002rq, Harvey:2005ab, Okuyama:2005gx, Kimura:2013zva}.

We describe (\ref{GLSM-KKM}) in the language of the bosonic component fields and consider it in the IR limit.
Here we introduce bosonic scalar component fields of $\Gamma$ as $\frac{1}{\sqrt{2}} (\gamma^3 + \i \gamma^4)$.
Using this expression, we obtain the duality relation among the component fields of $\Theta$ and $\Gamma$,
\begin{gather}
\frac{1}{g^2} r^3 \ = \ - \gamma^3
\, , \ls
\mp \frac{1}{g^2} (\del_0 \pm \del_1) \vartheta
\ = \ 
(D_0 \pm D_1) \gamma^4
\, , \label{DR-Hm-KKm}
\end{gather}
where the covariant derivative is defined as 
$D_m \gamma^4 = \del_m \gamma^4 + \sqrt{2} \sum_{a=1}^k A_{a,m}$.
We perform the following steps: 
integrate out all the auxiliary fields, 
and then restrict
the system only to the supersymmetric Higgs branch (\ref{Higgsbranch-ex1}).
Accordingly, we obtain
\begin{align}
\Scr{L}_{\text{KK}}
\ &= \ 
\sum_{a=1}^k \frac{1}{2 e_a^2} (F_{a,01})^2
- \frac{1}{2} \Big\{ \frac{1}{g^2} + \sum_{a=1}^k \frac{1}{\sqrt{2} R_a} \Big\} (\del_m \vec{r})^2 
\nn \\
\ & \ \ \ \ 
- \frac{g^2}{2} (D_m \gamma^4)^2
- \sum_{a=1}^k \sqrt{2} R_a \Big( 
A_{a,m} + \frac{1}{\sqrt{2}} {\omega}_{a,i} \del_m r^i 
\Big)^2
\nn \\
\ & \ \ \ \ 
+ \text{(total derivative terms)}
+ \text{(fermionic fields)}
\, . \label{GLSM-KKm-b1}
\end{align}
Here we re-labelled $\gamma^4 = \wt{\vartheta}$ to emphasize that this is the dual field of the original field $\vartheta$.
In order to consider the string sigma model, 
we take the IR limit $e_a \to \infty$.
Since the gauge fields are no longer dynamical in this limit, we integrate them out.
Plugging the solution into (\ref{GLSM-KKm-b1}), we finally obtain the NLSM for the $k$-centered KK-monopoles,
\begin{align}
\Scr{L}_{\text{KK}}
\ &= \ 
- \frac{1}{2} H (\del_m \vec{r})^2
- \half H^{-1} \Big( \del_m \wt{\vartheta} - \omega_{i} \, \del_m r^i \Big)^2
\nn \\
\ & \ \ \ \ 
+ \text{(total derivative terms)}
+ \text{(fermionic fields)}
\, . \label{GLSM-KKm-b2}
\end{align}
Again, the functions $\omega_i$ and $H$ defined in (\ref{omega-R}) and (\ref{GLSM-Hm-b3}) emerge.
The inverse of the harmonic function $H$ appears
in front of the second derivative term of $\wt{\vartheta}$.
There are no terms coupled to the antisymmetric tensor $\ve_{mn}$
which represents the target space B-field.
This is also a solution of ten-dimensional supergravity.
Indeed, the target space metric of this NLSM is the $k$-centered Taub-NUT space, i.e., a non-compact hyper-K\"{a}hler geometry.
In the absence of the target space B-field, we can immediately find that the target space dilaton becomes trivial.

\subsection{Exotic $5^2_2$-brane}
\label{subsect:522}

The exotic $5^2_2$-brane is constructed via the T-duality transformation along a transverse direction of the KK-monopoles.
However, there is only one isometry direction along $\vartheta$, which is the direction to connect between NS5-branes and KK-monopoles via the T-duality.
In order to make an additional isometry along a different direction,
we put an infinite number of KK-monopoles along the $r^2$-direction.
Smearing this direction, we can take the T-duality transformation along it.
This has been analyzed in the supergravity framework \cite{deBoer:2010ud, deBoer:2012ma}.
We also perform the same procedure in the framework of the GLSM \cite{Kimura:2013fda}.

We consider the duality transformation to the chiral superfield $\Psi$ in the GLSM for the KK-monopoles (\ref{GLSM-KKM}).
Since $\Psi$ contributes to the F-term as well as the D-terms,
we should apply the technique discussed in section \ref{subsect:DT-NCS-DF}.
Here we set the holomorphic function ${\cal W}$ to the identity.
The duality relation (\ref{PsiXi-NCS}) of the GLSM (\ref{GLSM-KKM}) is
\begin{align}
\frac{1}{g^2} (\Psi + \ol{\Psi})
\ &= \ 
- (\Xi + \ol{\Xi}) + \sqrt{2} \sum_{a=1}^k (C_a + \ol{C}{}_a) 
\, , \label{Psi-Gamma-522}
\end{align}
where $\Xi$ is the dual twisted chiral superfield whose scalar component fields are $\frac{1}{\sqrt{2}} (y^1 + \i y^2)$.
The dual GLSM is also obtained from the general result (\ref{DT-LDF-NCS}) 
after setting $\beta = 1$ \cite{Kimura:2013fda},
\begin{align}
\Scr{L}_{\text{E}} \ &= \ 
\sum_{a=1}^k \int \d^4 \theta \, \Big\{ 
\frac{1}{e_a^2} 
\Big( - |\Sigma_a|^2 + |\Phi_a|^2 \Big)
+ |Q_a|^2 \, \e^{-2 V_a} 
+ |\wt{Q}_a|^2 \, \e^{+2 V_a} 
\Big\}
\nn \\
\ & \ \ \ \ 
+ \int \d^4 \theta \, 
\frac{g^2}{2} \Big\{
- \Big( \Xi + \ol{\Xi} - \sqrt{2} \sum_{a=1}^k (C_a + \ol{C}{}_a) \Big)^2
+ \Big( \Gamma + \ol{\Gamma} + 2 \sum_{a=1}^k V \Big)^2
\Big\}
\nn \\
\ & \ \ \ \ 
+ \sum_{a=1}^k \Big\{
\sqrt{2} \int \d^2 \theta \, \big( \wt{Q}_a \Phi_a Q_a + s_a \, \Phi_a \big)
+ \text{(h.c.)} 
\Big\}
+ \sum_{a=1}^k \Big\{ 
\sqrt{2} \int \d^2 \wt{\theta} \, (t_a \, \Sigma_a)
+ \text{(h.c.)}
\Big\}
\nn \\
\ & \ \ \ \ 
- \sqrt{2} \int \d^4 \theta \, 
(\Psi - \ol{\Psi}) \sum_{a=1}^k (C_a - \ol{C}{}_a)
- \sqrt{2} \, \ve^{mn} \sum_{a=1}^k \del_m (\vartheta A_{a,n})
\, . \label{GLSM-522}
\end{align}
The key ingredient of this Lagrangian is the term containing the imaginary part of $\Psi$ in the fourth line.
Again we ignore the total derivative term including the gauge field $A_{a,m}$ in (\ref{GLSM-522}).

We investigate this dual GLSM in terms of the bosonic component fields.
First, the duality relation (\ref{Psi-Gamma-522}) is expanded in the following way,
\bsubeq \label{DR-KKm-E}
\begin{align}
\frac{1}{g^2} r^1 \ &= \ 
- y^1
+ \sum_{a=1}^k (\phi_{c,a} + \ol{\phi}{}_{c,a})
\, , \\
\frac{1}{g^2} (\del_0 + \del_1) r^2 \ &= \ 
- (\del_0 + \del_1) y^2 
+ \sum_{a=1}^k \big( B_{c\+,a} + \ol{B}{}_{c\+,a} \big)
\, , \\
\frac{1}{g^2} (\del_0 - \del_1) r^2 \ &= \ 
+ (\del_0 - \del_1) y^2 
+ \sum_{a=1}^k \big( A_{c=,a} + \ol{A}{}_{c=,a} \big)
\, .
\end{align}
\esubeq
In the same way as in the previous subsection,
we expand the GLSM (\ref{GLSM-522}) in terms of the bosonic component fields.
Since there exist many auxiliary fields,
we integrate them all out.
If we focus only on the supersymmetric Higgs branch,
we obtain the bosonic description of the GLSM (\ref{GLSM-522}) in the following form,
\begin{align}
\Scr{L}_{\text{E}}
\ &= \ 
\sum_{a=1}^k \frac{1}{2 e_a^2} (F_{a,01})^2
- \frac{1}{2 g^2} \Big\{ (\del_m r^1)^2 + (\del_m r^3)^2 \Big\}
- \frac{g^2}{2} \Big\{ (\del_m y^2)^2 + (D_m \gamma^4)^2 \Big\}
\nn \\
\ & \ \ \ \ 
- \sum_{a=1}^k \Big\{ 
|D_m q_a|^2
+ |D_m \wt{q}_a|^2
\Big\}
\nn \\
\ & \ \ \ \ 
+ \text{(total derivative terms)}
+ \text{(fermionic fields)}
\, . \label{GLSM-E-b1}
\end{align}
Notice that, in addition to the constraints (\ref{Higgsbranch-ex1}),
we have to impose the following constraint,
\begin{align}
0 \ &= \ 
\frac{g^2}{2} \sum_{a,b} 
(A_{c=,a} + \ol{A}{}_{c=,a}) (B_{c\+,b} + \ol{B}{}_{c\+,b})
\nn \\
\ &= \ 
- \frac{1}{2 g^2} (\del_m r^2)^2
+ \frac{g^2}{2} (\del_m y^2)^2 
+ \ve^{mn} (\del_m r^2) (\del_n y^2)
\, . \label{add-HB-522}
\end{align}
Indeed the first line in the right-hand side appears in the Lagrangian as a potential term, 
which should vanish to preserve supersymmetry (for the detail, see \cite{Kimura:2013fda}).
The first line is converted to the second line via (\ref{DR-KKm-E}).
Surprisingly, if we substitute this constraint into (\ref{GLSM-E-b1}),
the second derivative of the dual scalar field $y^2$ disappears, whilst the second derivative of the original scalar field $r^2$ is recovered. 
In addition, a topological term is generated.
Indeed this process plays a central role in the T-duality transformation 
later in the discussion.

We investigate the IR limit $e_a \to \infty$ in order to construct the string worldsheet sigma model.
In this limit the kinetic terms of the gauge fields disappear and all the gauge fields become auxiliary fields.
Integrating them out, we obtain 
\begin{align}
\Scr{L}_{\text{E}}
\ &= \ 
- \frac{1}{2} H \Big\{ (\del_m r^1)^2 + (\del_m r^2)^2 + (\del_m r^3)^2 \Big\}
+ \ve^{mn} (\del_m r^2) (\del_n y^2) 
\nn \\
\ & \ \ \ \ 
- \frac{1}{2} H^{-1} (\del_m \wt{\vartheta})^2 
- \half ({\omega}_{2})^2 H^{-1} (\del_m r^2)^2
+ {\omega}_{2} H^{-1} 
(\del_m \wt{\vartheta}) (\del^m r^2) 
\nn \\
\ & \ \ \ \ 
- \half ({\omega}_{1})^2 H^{-1} (\del_m r^1)^2
- {\omega}_{1} {\omega}_{2} H^{-1} (\del_m r^1) (\del^m r^2)
+ {\omega}_{1} H^{-1} (\del_m \wt{\vartheta}) (\del^m r^1)  
\nn \\
\ & \ \ \ \ 
+ \text{(total derivative terms)}
+ \text{(fermionic fields)}
\, . \label{GLSM-E-b2}
\end{align}
The functions $\omega_i$ and $H$
are defined in (\ref{omega-R}) and (\ref{GLSM-Hm-b3}).
We should notice that this is not the final form of the NLSM 
because the functions $H$ and $\omega_i$ still depend on $r^2$.
This implies that there is no isometry along this direction.
Here we use a trick to generate an isometry.
As mentioned before, 
we prepare an infinite number of the KK-monopoles in the framework of supergravity. 
In the GLSM framework, we prepare an infinite number of vector multiplets, i.e., we take the infinity limit $k \to \infty$.
In this limit, the functions $H$ and $\omega_i$ do not depend on $r^2$ any more,
\bsubeq \label{Omega-kinfty-E}
\begin{gather}
H \ \xrightarrow{k\to\infty} \ 
h_0 + \sigma \log \frac{\mu}{\varrho}
\, , \ls
\sigma \ = \ \frac{1}{\sqrt{2} \, \pi {\cal R}_2}
\, ,
\\
\omega_1 \ \xrightarrow{k\to\infty} \ 0
\, , \ls
\omega_2 \ \xrightarrow{k\to\infty} \ 
\omega_{\varrho} 
\ = \ 
\sigma \arctan \Big( \frac{r^3}{r^1} \Big)
\, ,
\end{gather}
\esubeq
where $\varrho^2 = (r^1)^2 + (r^3)^2$, and ${\cal R}_2$ is the size of the $r^2$-direction.
Finally we evaluate the equation of motion for the scalar field $r^2$.
This is a non-dynamical field after the duality transformation (\ref{DR-KKm-E}), even though the second derivative term appears in the constraint (\ref{add-HB-522}).
Plugging its solution of the equation of motion into (\ref{GLSM-E-b2}) with the limit (\ref{Omega-kinfty-E}), we consequently obtain 
\begin{align}
\Scr{L}_{\text{E}}
\ &= \ 
- \frac{1}{2} H \Big\{ (\del_m r^1)^2 + (\del_m r^3)^2 \Big\}
- \half H K^{-1} \Big\{ (\del_m y^2)^2 + (\del_m \wt{\vartheta})^2 \Big\}
\nn \\
\ & \ \ \ \ 
- \omega_{\varrho} K^{-1} \, \ve^{mn} (\del_m y^2) (\del_n \wt{\vartheta})
\nn \\
\ & \ \ \ \ 
+ \text{(total derivative terms)}
+ \text{(fermionic fields)}
\, . \label{GLSM-E-b3}
\end{align}
The kinetic term of $y^2$ is recovered due to the existence of the topological term.
The target space represents nothing but the exotic $5^2_2$-brane with B-field in ten-dimensional supergravity \cite{deBoer:2010ud, deBoer:2012ma}.
We remark that this configuration is derived from that of the KK-monopoles via the Buscher rule.
We conclude that the duality transformation rule in the presence of the F-term discussed in section \ref{subsect:DT-NCS-DF} is consistent with the T-duality transformation rule,
and confirmed that 
the existence of the F-term and its contribution to the duality transformation rule in GLSMs are necessary
to reproduce the correct T-duality transformations.

\section{GLSM for $A_1$-type ALE space and its T-duality}
\label{sect:TA-ALE}

In this section we investigate the GLSM for the $A_1$-type ALE space \cite{Eguchi:1978gw} and its T-duality.
As discussed in our previous work \cite{Kimura:2014bxa},
the F-term plays a central role in constructing
the correct target space configuration.
This should also be true in the T-duality transformation procedure.
In section \ref{sect:GLSMF-ALE} we evaluate the GLSM for the $A_1$-type ALE space.
In section \ref{sect:GLSMF-TALE} we investigate the T-duality transformation of this GLSM by virtue of the technique discussed in section \ref{subsect:DT-CCS-DF}.

\subsection{GLSM for $A_1$-type ALE space}
\label{sect:GLSMF-ALE}

The $A$-type ALE space can be represented as a toric variety.
This indicates that the GLSM formulation is useful to discuss the string worldsheet sigma model on that space.
An explicit form of the GLSM for the $A_1$-type ALE space is described in terms of two $\N=(4,4)$ vector multiplets 
$\{ V, \Phi \}$,
$\{ \wt{V}, \wt{\Phi} \}$,
and three $\N=(4,4)$ charged hypermultiplets 
$\{ A_1, B_1 \}$,
$\{ A_2, B_2 \}$,
$\{ A_3, B_3 \}$, and 
the complexified FI parameter $t = \frac{1}{\sqrt{2}} (t_1 + \i t_2)$ \cite{Kimura:2014bxa},
\begin{align}
\Scr{L}_{\text{EH}} \ &= \ 
\int \d^4 \theta \, \Big\{
\frac{1}{e^2} \Big( 
- |\Sigma|^2 + |\Phi|^2
\Big)
+ \frac{1}{\wt{e}^2} \Big( 
- |\wt{\Sigma}|^2 + |\wt{\Phi}|^2
\Big)
\Big\}
\nn \\
\ & \ \ \ \ 
+ \int \d^4 \theta \, \Big\{
  |A_1|^2 \, \e^{+2V}
+ |A_2|^2 \, \e^{-4 V - 2 \alpha \wt{V}}
+ |A_3|^2 \, \e^{+2V}
\Big\}
\nn \\
\ & \ \ \ \ 
+ \int \d^4 \theta \, \Big\{ 
  |B_1|^2 \, \e^{-2V}
+ |B_2|^2 \, \e^{+4 V + 2 \alpha \wt{V}}
+ |B_3|^2 \, \e^{-2V}
\Big\}
\nn \\
\ & \ \ \ \ 
+ \Big\{ \sqrt{2} \int \d^2 \theta \, 
\Big(
\Phi \big(
- A_1 B_1 + 2 A_2 B_2 - A_3 B_3
\big)
+ \alpha \, \wt{\Phi} A_2 B_2
\Big)
+ \text{(h.c.)}
\Big\}
\nn \\
\ & \ \ \ \ 
+ \Big\{ \sqrt{2} \int \d^2 \wt{\theta} \, 
(- t \, \Sigma)
+ \text{(h.c.)}
\Big\}
\, . \label{GLSMF_A1-ALE}
\end{align}
The coefficients in the F-term are determined by the charges of the hypermultiplets needed to preserve $SU(2)_R$ symmetry.
In the following discussions, 
we derive the low energy supersymmetric NLSM in two ways: 
one is the NLSM in the superfield formalism
where the K\"{a}hler potential of the system is explicitly 
described,
the other is the NLSM formulated by the component fields.
Comparing to the string worldsheet sigma model (\ref{string-NLSM-bosons}),
we can read off the target space configuration.

\subsubsection{Analysis by superfields}

We analyze the IR limit of the GLSM (\ref{GLSMF_A1-ALE}) in the superfield formalism.
In the IR limit $e, \wt{e} \to \infty$, 
all the vector multiplets become auxiliary fields,
so we can integrate them out.
Afterward, their field equations are  
\bsubeq \label{EOM-ASF}
\begin{align}
0 \ &= \ 
(|A_1|^2 + |A_3|^2) \, \e^{+ 2 V}
- (|B_1|^2 + |B_3|^2) \, \e^{-2 V}
- \sqrt{2} \, t_1
\nn \\
\ & \ \ \ \ 
- 2 |A_2|^2 \, \e^{- 4V-2 \alpha \wt{V}}
+ 2 |B_2|^2 \, \e^{+ 4 V+2 \alpha \wt{V}}
\, , \label{EOM-V} \\
0 \ &= \ 
- \alpha |A_2|^2 \, \e^{-4V-2 \alpha \wt{V}}
+ \alpha |B_2|^2 \, \e^{+ 4V + 2 \alpha \wt{V}}
\, , \label{EOM-V'} \\
0 \ &= \ 
- A_1 B_1 + 2 A_2 B_2 - A_3 B_3 
\, , \label{EOM-Phi} \\
0 \ &= \ 
\alpha A_2 B_2 
\, . \label{EOM-Phi'}
\end{align}
\esubeq
Immediately we find that the hypermultiplet $\{ A_2, B_2 \}$ 
and the vector multiplet $\{ \wt{V} , \wt{\Phi} \}$ vanish.
The solution of the field equation for the vector multiplet $\{ V, \Phi \}$ is
\bsubeq \label{EOM-VPhi}
\begin{align}
\e^{+ 2 V} \ &= \ 
\frac{1}{2 M_1} \Big\{ \sqrt{2} \, t_1 + \sqrt{(\sqrt{2} \, t_1)^2 + 4 M_1 M_2} \Big\}
\, , \\
M_1 \ &\equiv \ 
|A_1|^2 + |A_3|^2
\, , \ls
M_2 \ \equiv \ 
|B_1|^2 + |B_3|^2
\, .
\end{align}
\esubeq
Plugging this into (\ref{GLSMF_A1-ALE}) under the IR limit, 
we obtain the NLSM in the superfield formalism,
\begin{align}
\Scr{L}
\ &= \ 
\int \d^4 \theta \, \Bigg\{
\sqrt{(\sqrt{2} \, t_1)^2 + 4 M_1 M_2} 
- \sqrt{2} \, t_1 \log \frac{\sqrt{2} \, t_1 + \sqrt{(\sqrt{2} \, t_1)^2 + 4 M_1 M_2}}{2 M_1}
\Bigg\}
\nn \\
\ & \ \ \ \ 
- \sqrt{2} \, t_2 F_{01}
\, . \label{GLSMF44-IR2}
\end{align}
This is the K\"{a}hler potential of the $A_1$-type ALE space.
The term $t_2 F_{01}$ will play a central role in the quantum instanton corrections to this system, though we do not discuss it in this work.
Of course it is possible to expand (\ref{GLSMF44-IR2})
in terms of the component fields.
We parameterize the scalar component fields of the dynamical hypermultiplets 
$\{ A_1, B_1 \}$ and $\{ A_3, B_3 \}$,
setting the FI parameter to $\sqrt{2} \, t_1 = \a^2/2$,
\bsubeq \label{A13B13-NLSM}
\begin{alignat}{2}
a_1 
\ &= \ 
\frac{r}{2} \cos \frac{\vartheta}{2} \, \e^{\frac{\i}{2} (\wt{a} \psi + \varphi)}
\, , &\ls
a_3
\ &= \ 
\frac{r}{2} \sin \frac{\vartheta}{2} \, \e^{\frac{\i}{2} (\wt{a} \psi - \varphi)}
\, , \\
b_1 
\ &= \ 
\frac{r}{2} \sin \frac{\vartheta}{2} \, \e^{\frac{\i}{2} (\wt{c} \psi - \varphi)}
\, , &\ls
b_3
\ &= \ 
- \frac{r}{2} \cos \frac{\vartheta}{2} \, \e^{\frac{\i}{2} (\wt{c} \psi + \varphi)}
\, .
\end{alignat}
\esubeq
Here 
the scalar field $r$ represents the radial coordinate on the target space of the sigma model, 
while the scalar fields $\{ \vartheta, \varphi, \psi \}$ indicate the Euler angles 
$\vartheta \in [0, \pi)$, $\varphi \in [ 0, 2 \pi)$, and $\psi \in [0, 4 \pi)$.
The constants $\{ \wt{a}, \wt{c} \}$ are arbitrary under the constraint $\wt{a} + \wt{c} = + 2$.
Note that the normalization is slightly different from that of \cite{Kimura:2014bxa} in order to obtain the canonical form of the NLSM.
Introducing a new variable $\rho^4 = \a^4 + r^4$, 
we find that the Lagrangian (\ref{GLSMF44-IR2}) in the component field formulation is given as
\begin{align}
\Scr{L}
\ &= \
- \half \Big( 1 - \frac{\a^4}{\rho^4} \Big)^{-1} (\del_m \rho)^2
- \frac{\rho^2}{8} \Big\{ (\del_m \vartheta)^2 + (\del_m \varphi)^2 \sin^2 \vartheta \Big\}
- \Big( 1 - \frac{\a^4}{\rho^4} \Big)
\frac{\rho^2}{8} \Big\{ (\del_m \psi) + (\del_m \varphi) \cos \vartheta \Big\}^2
\nn \\
\ & \ \ \ \ 
- \sqrt{2} \, t_2 F_{01} 
+ \text{(fermionic fields)}
\, . \label{GLSMF44-IR2-2}
\end{align}
The constants $\{ \wt{a}, \wt{c} \}$ do not appear in any terms.
Compared with the string sigma model Lagrangian (\ref{string-NLSM-bosons}),
we find that the target space geometry of (\ref{GLSMF44-IR2-2}) represents the 
$A_1$-type ALE space.
Next we will derive the same NLSM in a different way.

\subsubsection{Analysis by component fields}

We derive the NLSM (\ref{GLSMF44-IR2-2}) in a different way.
First, we expand the GLSM (\ref{GLSMF_A1-ALE}) in terms of the component fields,
\begin{align}
\Scr{L} \ &= \ 
\frac{1}{e^2} \Big\{
\half (F_{01})^2 
- |\del_m \sigma|^2
- |\del_m \phi|^2
\Big\}
+ \frac{1}{\wt{e}^2} \Big\{
\half (\wt{F}_{01})^2 
- |\del_m \wt{\sigma}|^2
- |\del_m \wt{\phi}|^2
\Big\}
- \sqrt{2} \, t_2 F_{01}
\nn \\
\ & \ \ \ \ 
- |D_m a_1|^2
- |D_m a_2|^2
- |D_m a_3|^2
- |D_m b_1|^2
- |D_m b_2|^2
- |D_m b_3|^2
\nn \\
\ & \ \ \ \  
- 2 |\sigma|^2 \Big\{ (|a_1|^2 + |a_3|^2) + (|b_1|^2 + |b_3|^2) \Big\}
- 2 \big| 2 \sigma + \alpha \wt{\sigma} \big|^2 (|a_2|^2 + |b_2|^2) 
\nn \\
\ & \ \ \ \  
+ \frac{1}{2 e^2} (D_{V})^2
+ \Big\{ (|a_1|^2 + |a_3|^2)
- (|b_1|^2 + |b_3|^2)
- 2 (|a_2|^2 - |b_2|^2) 
- \sqrt{2} \, t_1 
\Big\} D_V
\nn \\
\ & \ \ \ \  
+ \frac{1}{2 \wt{e}^2} (D_{\wt{V}})^2
- \alpha (|a_2|^2 - |b_2|^2) D_{\wt{V}} 
\nn \\
\ & \ \ \ \  
+ \frac{1}{e^2} |D_{\Phi}|^2
- \i \sqrt{2} \Big\{ a_1 b_1 + a_3 b_3 - 2 a_2 b_2 \Big\} D_{\Phi}
+ \i \sqrt{2} \Big\{ \ol{a}{}_1 \ol{b}{}_1 + \ol{a}{}_3 \ol{b}{}_3 - 2 \ol{a}{}_2 \ol{b}{}_2 \Big\} \ol{D}{}_{\Phi}
\nn \\
\ & \ \ \ \  
+ \frac{1}{\wt{e}^2} |D_{\wt{\Phi}}|^2
+ \i \sqrt{2} \, \alpha \, a_2 b_2 \, D_{\wt{\Phi}} 
- \i \sqrt{2} \, \alpha \,  \ol{a}{}_2 \ol{b}{}_2 \, \ol{D}{}_{\wt{\Phi}}
\nn \\
\ & \ \ \ \  
+ |F_1|^2
+ |\wt{F}_1|^2
- \i \sqrt{2} \Big\{
\phi \big( a_1 \wt{F}_1 + b_1 F_1 \big)
\Big\}
+ \i \sqrt{2} \Big\{
\ol{\phi} \big( \ol{a}{}_1 \ol{\wt{F}}{}_1 + \ol{b}{}_1 \ol{F}{}_1 \big)
\Big\}
\nn \\
\ & \ \ \ \  
+ |F_2|^2
+ |\wt{F}_2|^2
+ \i \sqrt{2} \, \Big\{
(2 \phi + \alpha \wt{\phi}) \big( a_2 \wt{F}_2 + b_2 F_2 \big)
\Big\}
- \i \sqrt{2} \, \Big\{
(2 \ol{\phi} + \alpha \ol{\wt{\phi}}) \big( \ol{a}{}_2 \ol{\wt{F}}{}_2 + \ol{b}{}_2 \ol{F}{}_2 \big)
\Big\}
\nn \\
\ & \ \ \ \  
+ |F_3|^2
+ |\wt{F}_3|^2
- \i \sqrt{2} \Big\{
\phi \big( a_3 \wt{F}_3 + b_3 F_3 \big)
\Big\}
+ \i \sqrt{2} \Big\{
\ol{\phi} \big( \ol{a}{}_3 \ol{\wt{F}}{}_3 + \ol{b}{}_3 \ol{F}{}_3 \big)
\Big\}
\nn \\
\ & \ \ \ \ 
+ \text{(fermionic fields)}
\, , \label{GLSMF-b1}
\end{align}
where the covariant derivatives for the charged scalar fields are defined as 
\bsubeq
\begin{alignat}{2}
D_m a_1 \ &= \ 
\del_m a_1 + \i A_m \, a_1
\, , &\ls
D_m b_1 \ &= \ 
\del_m b_1 - \i A_m \, b_1
\, , \\
D_m a_2 \ &= \ 
\del_m a_2 - 2 \i A_m \, a_2
- \i \alpha \, \wt{A}_m \, a_2
\, , &\ls
D_m b_2 \ &= \ 
\del_m b_2 + 2 \i A_m \, b_2
+ \i \alpha \, \wt{A}_m \, b_2
\, , \\
D_m a_3 \ &= \ 
\del_m a_3 + \i A_m \, a_3
\, , &\ls
D_m b_3 \ &= \ 
\del_m b_3 - \i A_m \, b_3
\, .
\end{alignat}
\esubeq
Second, we integrate out the auxiliary fields $\{ D_V, D_{\wt{V}}, D_{\Phi}, D_{\wt{\Phi}}, F_i, \wt{F}_i \}$.
Their field equations are 
\bsubeq \label{EOM-aux-GLSMF}
\begin{gather}
\begin{align}
0 \ &= \ 
\frac{1}{e^2} D_V
+ \Big\{ (|a_1|^2 + |a_3|^2)
- (|b_1|^2 + |b_3|^2)
- 2 (|a_2|^2 - |b_2|^2) 
- \sqrt{2} \, t_1 
\Big\}
\, , \\
0 \ &= \ 
\frac{1}{\wt{e}^2} D_{\wt{V}}
- \alpha (|a_2|^2 - |b_2|^2)
\, , \\
0 \ &= \
\frac{1}{e^2} \ol{D}{}_{\Phi} 
- \i \sqrt{2} \Big\{ a_1 b_1 + a_3 b_3 - 2 a_2 b_2 \Big\}
\, , \\
0 \ &= \ 
\frac{1}{\wt{e}^2} \ol{D}{}_{\wt{\Phi}}
+ \i \sqrt{2} \, \alpha \, a_2 b_2 
\, , 
\end{align}
\\
\begin{alignat}{2}
0 \ &= \ 
\ol{F}{}_1 - \i \sqrt{2} \, \phi \, b_1
\, , &\ls
0 \ &= \ 
\ol{\wt{F}}{}_1 - \i \sqrt{2} \, \phi \, a_1
\, , \\
0 \ &= \ 
\ol{F}{}_2 + \i \sqrt{2} \, (2 \phi + \alpha \wt{\phi}) \, b_2
\, , &\ls
0 \ &= \ 
\ol{\wt{F}}{}_2 + \i \sqrt{2} \, (2 \phi + \alpha \wt{\phi}) \, a_2
\, , \\
0 \ &= \ 
\ol{F}{}_3 - \i \sqrt{2} \, \phi \, b_3
\, , &\ls
0 \ &= \ 
\ol{\wt{F}}{}_3 - \i \sqrt{2} \, \phi \, a_3
\, .
\end{alignat}
\end{gather}
\esubeq
Plugging them into (\ref{GLSMF-b1}), 
we obtain 
\begin{align}
\Scr{L} \ &= \ 
\frac{1}{e^2} \Big\{
\half (F_{01})^2 
- |\del_m \sigma|^2
- |\del_m \phi|^2
\Big\}
+ \frac{1}{\wt{e}^2} \Big\{
\half (\wt{F}_{01})^2 
- |\del_m \wt{\sigma}|^2
- |\del_m \wt{\phi}|^2
\Big\}
- \sqrt{2} \, t_2 F_{01}
\nn \\
\ & \ \ \ \ 
- |D_m a_1|^2
- |D_m a_2|^2
- |D_m a_3|^2
- |D_m b_1|^2
- |D_m b_2|^2
- |D_m b_3|^2
\nn \\
\ & \ \ \ \  
- 2 (|\sigma|^2 + |\phi|^2) 
\Big\{ (|a_1|^2 + |a_3|^2) + (|b_1|^2 + |b_3|^2) \Big\}
\nn \\
\ & \ \ \ \  
- 2 \Big\{ \big| 2 \sigma + \alpha \wt{\sigma} \big|^2 
+ \big| 2 \phi + \alpha \wt{\phi} \big|^2 \Big\} (|a_2|^2 + |b_2|^2)  
\nn \\
\ & \ \ \ \  
- \frac{e^2}{2} \Big\{
(|a_1|^2 + |a_3|^2)
- (|b_1|^2 + |b_3|^2)
- 2 (|a_2|^2 - |b_2|^2) 
- \sqrt{2} \, t_1 
\Big\}^2
- \frac{\alpha^2 \wt{e}^2}{2} (|a_2|^2 - |b_2|^2)^2
\nn \\
\ & \ \ \ \  
- 2 e^2 \big| a_1 b_1 + a_3 b_3 - 2 a_2 b_2 \big|^2
- 2 \alpha^2 \wt{e}^2 |a_2|^2 |b_2|^2
\nn \\
\ & \ \ \ \ 
+ \text{(fermionic fields)}
\, . \label{GLSMF-b2}
\end{align}
We are interested in the supersymmetric Higgs branch in this system,
so we impose the constraints
\bsubeq \label{HB-ALE}
\begin{align}
0 \ &= \ 
\sigma \ = \ \phi
\, , \ls
0 \ = \ 
\wt{\sigma} \ = \ \wt{\phi}
\, , \\
0 \ &= \ 
(|a_1|^2 + |a_3|^2)
- (|b_1|^2 + |b_3|^2)
- 2 (|a_2|^2 - |b_2|^2) 
- \sqrt{2} \, t_1 
\, , \label{D-flat-comp} \\
0 \ &= \ 
a_1 b_1 + a_3 b_3 - 2 a_2 b_2
\, , \\
0 \ &= \ 
|a_2|^2 - |b_2|^2
\, , \ls
0 \ = \ 
|a_2|^2 |b_2|^2
\, .
\end{align}
\esubeq
We immediately find that the scalar fields $\{ a_2, b_2 \}$ vanish.
The Lagrangian and the constraints in the Higgs branch are reduced to 
\bsubeq \label{GLSMF-b3}
\begin{align}
\Scr{L}
\ &= \ 
\frac{1}{2 e^2} (F_{01})^2 
+ \frac{1}{2 \wt{e}^2} (\wt{F}_{01})^2 
- |D_m a_1|^2
- |D_m a_3|^2
- |D_m b_1|^2
- |D_m b_3|^2
\nn \\
\ & \ \ \ \ 
- \sqrt{2} \, t_2 F_{01}
+ \text{(fermionic fields)}
\, , \\
\sqrt{2} \, t_1
\ &= \ \frac{\a^2}{2} 
\ = \ 
(|a_1|^2 + |a_3|^2)
- (|b_1|^2 + |b_3|^2)
\, , \ls
0 \ = \ 
a_1 b_1 + a_3 b_3
\, . \label{constr-GLSMF-b}
\end{align}
\esubeq
Now we take the IR limit $e, \wt{e} \to \infty$.
Since the gauge field $\wt{A}_m$ is decoupled from the system and $A_m$ is no longer dynamical,
we integrate them out.
The solution is 
\bsubeq
\begin{align}
A_m \ &= \ 
- \frac{\i}{2 \Z} 
\Big\{
  (a_1 \del_m \ol{a}{}_1 - \ol{a}{}_1 \del_m a_1)
+ (a_3 \del_m \ol{a}{}_3 - \ol{a}{}_3 \del_m a_3)
\Big\}
\nn \\
\ & \ \ \ \ 
+ \frac{\i}{2 \Z} 
\Big\{
  (b_1 \del_m \ol{b}{}_1 - \ol{b}{}_1 \del_m b_1)
+ (b_3 \del_m \ol{b}{}_3 - \ol{b}{}_3 \del_m b_3)
\Big\}
\, , \\
\Z \ &\equiv \ 
(|a_1|^2 + |a_3|^2) + (|b_1|^2 + |b_3|^2) 
\, .
\end{align}
\esubeq
Substituting this into the Lagrangian (\ref{GLSMF-b3}) in the IR limit, 
we find the NLSM Lagrangian,
\begin{align}
\Scr{L}
\ &= \ 
- |\del_m a_1|^2
- |\del_m a_3|^2
- |\del_m b_1|^2
- |\del_m b_3|^2
\nn \\
\ & \ \ \ \ 
- \frac{1}{4 \Z} \Big\{
(a_1 \del_m \ol{a}{}_1 - \ol{a}{}_1 \del_m a_1)
+ (a_3 \del_m \ol{a}{}_3 - \ol{a}{}_3 \del_m a_3)
- (b_1 \del_m \ol{b}{}_1 - \ol{b}{}_1 \del_m b_1)
- (b_3 \del_m \ol{b}{}_3 - \ol{b}{}_3 \del_m b_3)
\Big\}^2
\nn \\
\ & \ \ \ \ 
- \sqrt{2} \, t_2 F_{01}
+ \text{(fermionic fields)}
\, . \label{GLSMF-b4}
\end{align}
In order to approach the description (\ref{GLSMF44-IR2-2}) under the constraints (\ref{constr-GLSMF-b}),
we introduce the following notation,
\bsubeq \label{A13B13-GLSM}
\begin{alignat}{2}
a_1 \ &= \ 
\frac{\sqrt{\rho^2 + \a^2}}{2} \, 
\cos \frac{\vartheta}{2} \, \e^{\frac{\i}{2} (\wt{a} \psi + \varphi)}
\, , &\ls
a_3 \ &= \ 
\frac{\sqrt{\rho^2 + \a^2}}{2} \, 
\sin \frac{\vartheta}{2} \, \e^{\frac{\i}{2} (\wt{a} \psi - \varphi)}
\, , \\
b_1 \ &= \ 
\frac{\sqrt{\rho^2 - \a^2}}{2} \, 
\sin \frac{\vartheta}{2} \, \e^{\frac{\i}{2} (\wt{c} \psi -\varphi)}
\, , &\ls
b_3 \ &= \ 
- \frac{\sqrt{\rho^2 - \a^2}}{2} \, 
\cos \frac{\vartheta}{2} \, \e^{\frac{\i}{2} (\wt{c} \psi + \varphi)}
\, .
\end{alignat}
\esubeq
Here the constants $\{ \wt{a}, \wt{c} \}$ are arbitrary under the constraint $\wt{a} + \wt{c} = + 2$.
Notice that the scale factors of the fields $\{ a_i, b_i \}$ in (\ref{A13B13-GLSM}) are slightly different from those in (\ref{A13B13-NLSM}):
the former respects the equation of motion for the auxiliary fields (\ref{EOM-aux-GLSMF})
and the latter comes from the field equations for vector superfields (\ref{EOM-ASF}).
Both (\ref{EOM-aux-GLSMF}) and (\ref{EOM-ASF}) provide the D-term condition.
The difference in the scale factors is the contributions of the gauge field and derivative terms in (\ref{EOM-ASF}), 
which does not appear in (\ref{EOM-aux-GLSMF}).
Substituting (\ref{A13B13-GLSM}) into (\ref{GLSMF-b4}), 
we obtain exactly the same Lagrangian (\ref{GLSMF44-IR2-2}).

\subsection{T-duality of GLSM for $A_1$-type ALE space}
\label{sect:GLSMF-TALE}

In this subsection we investigate the duality transformation of the GLSM (\ref{GLSMF_A1-ALE}) by virtue of the new technique discussed in section \ref{subsect:DT-CCS-DF}.
We dualize the chiral superfield $A_1$ to the twisted chiral superfield $Y_1$.
In this procedure we set the holomorphic function ${\cal W}$ in (\ref{LDF-CCS}) to the chiral superfield $B_1$, the partner of $A_1$ in the $SU(2)_R$ doublet.
We should also set the constant $\wt{\alpha} = \alpha$ in (\ref{LDF-CCS}) to $+1$, which is the $U(1)$ charge of $A_1$ in accordance with the gauge symmetry of $V$.
The duality relation associated with (\ref{AY-dual-CCS}) is given as
\bsubeq \label{A1Y1_A1-ALE}
\begin{align}
|A_1|^2 \, \e^{+ 2V}
\ &= \ 
\frac{\F_1^2}{4}
\ = \ 
\beta_1 (Y_1 + \ol{Y}{}_1)
+ \frac{\T_1^2}{2} 
- \frac{\T_1}{2} \sqrt{\T_1^2 + 4 \beta_1 (Y_1 + \ol{Y}{}_1)}
\, , \\
A_1 \ &= \ \e^{\Psi_1}
\, ,
\end{align}
\esubeq
where we introduced the following functions,
\bsubeq \label{FT-A1}
\begin{align}
\F_1 \ &= \ 
- \T_1 + \sqrt{\T_1^2 + 4 \beta_1 (Y_1 + \ol{Y}{}_1)}
\, , \\
\T_1 \ &= \ 
- \sqrt{2} \, \e^{- V} \Big\{
\e^{\half (\Psi_1 - \ol{\Psi}{}_1)} B_1 C
+ \e^{- \half (\Psi_1 - \ol{\Psi}{}_1)} \ol{B}{}_1 \ol{C}
\Big\}
\, , \\
\Phi \ &= \ 
\ol{D}{}_+ \ol{D}{}_- C
\, .
\end{align}
\esubeq
Applying the general result (\ref{DT-LDF-CCS}) to (\ref{GLSMF_A1-ALE}), we obtain the dual GLSM for the $A_1$-type ALE space,
\begin{align}
\Scr{L} \ &= \ 
\int \d^4 \theta \, \Big\{
\frac{1}{e^2} \Big( 
- |\Sigma|^2 + |\Phi|^2
\Big)
+ \frac{1}{\wt{e}^2} \Big( 
- |\wt{\Sigma}|^2 + |\wt{\Phi}|^2
\Big)
\Big\}
\nn \\
\ & \ \ \ \ 
+ \int \d^4 \theta \, \Big\{
|A_3|^2 \, \e^{+2V}
+ (|B_1|^2 + |B_3|^2) \, \e^{-2V}
+ |A_2|^2 \, \e^{-4 V - 2 \alpha \wt{V}}
+ |B_2|^2 \, \e^{+4 V + 2 \alpha \wt{V}}
\Big\}
\nn \\
\ & \ \ \ \ 
+ \int \d^4 \theta \, \Big\{
- 2 \beta_1 (Y_1 + \ol{Y}{}_1) \log \F_1
+ \half \F_1 \T_1
\Big\}
\nn \\
\ & \ \ \ \ 
+ \int \d^4 \theta \, \Big\{
- 2 \sqrt{2} \, (A_3 B_3 - 2 A_2 B_2) C
- 2 \sqrt{2} \, (\ol{A}{}_3 \ol{B}{}_3 - 2 \ol{A}{}_2 \ol{B}{}_2) \ol{C}
\Big\}
\nn \\
\ & \ \ \ \ 
+ \Big\{
\sqrt{2} \int \d^2 \theta \, \alpha \, \wt{\Phi} A_2 B_2
+ \text{(h.c.)}
\Big\}
+ \Big\{
\sqrt{2} \int \d^2 \wt{\theta} \, 
(\beta_1 Y_1 - t) \Sigma
+ \text{(h.c.)}
\Big\}
\, . \label{DT-GLSMF_A1-ALE}
\end{align}
We will determine the constant $\beta_1$ later.
In the following discussions,
we will derive the K\"{a}hler potential of the dualized system. 
Further we will construct the NLSM in terms of the component fields.
We will find that the target space configuration of the NLSM denotes the T-dualized geometry with a non-vanishing B-field in the correct way.

\subsubsection{Analysis by superfields}

First we investigate the low energy NLSM in the superfield formalism.
We note that the system involves $C$ rather than its original form $\Phi$.
The field equations for the vector multiplets $\{ V, C \}$ and $\{ \wt{V}, \wt{\Phi} \}$ in the IR limit $e, \wt{e} \to \infty$ are 
\bsubeq
\begin{align}
0 \ &= \ 
|A_3|^2 \, \e^{2 V}
- (|B_1|^2 + |B_3|^2) \, \e^{-2 V}
+ \frac{\F_1^2}{4}
- \sqrt{2} \, t_1
\nn \\
\ & \ \ \ \ 
- 2 |A_2|^2 \, \e^{- 4 V - 2 \alpha \wt{V}}
+ 2 |B_2|^2 \, \e^{4 V + 2 \alpha \wt{V}}
\, , \label{EOM-V-TALE} \\
0 \ &= \ 
- \sqrt{2} \, \e^{-V + \half (\Psi_1 - \ol{\Psi}{}_1)} B_1 \F_1
- 2 \sqrt{2} \, (A_3 B_3 - 2 A_2 B_2)
\, , \label{EOM-C-TALE} \\
0 \ &= \ 
- \alpha |A_2|^2 \, \e^{- 4 V - 2 \alpha \wt{V}}
+ \alpha |B_2|^2 \, \e^{4 V + 2 \alpha \wt{V}}
\, , \label{EOM-Vt-TALE} \\
0 \ &= \ A_2 B_2
\, . \label{EOM-Phit-TALE}
\end{align}
\esubeq
Again,
we immediately find that the supermultiplet $\{ A_2, B_2 \}$ vanishes,
and the vector multiplet $\{ \wt{V}, \wt{\Phi} \}$ is decoupled from the system.
Multiplying $C$ to equation (\ref{EOM-C-TALE}) and 
adding its hermitian conjugate we obtain
\begin{align}
0 \ &= \ 
\F_1 \T_1 - 2 \sqrt{2} \, (A_3 B_3 C + \ol{A}{}_3 \ol{B}{}_3 \ol{C})
\, . \label{EOM-C-Dual1-2'} 
\end{align}
In terms of the definition of the functions $\{ \F_1, \T_1 \}$,
we obtain 
\begin{gather}
\F_1^2 
\ = \ 
- 4 \sqrt{2} \, (A_3 B_3 C + \ol{A}{}_3 \ol{B}{}_3 \ol{C})
+ 4 \beta_1 (Y_1 + \ol{Y}{}_1)
\, . \label{F12-TALE}
\end{gather}
We also obtain the following form from the equation (\ref{EOM-C-TALE}),
\begin{align}
|A_3|^2 \, \e^{2 V} 
\ &= \ 
\frac{|B_1|^2}{|B_3|^2} \frac{\F_1^2}{4}
\, . \label{A3VB1B3-F1}
\end{align}
Plugging (\ref{F12-TALE}) and (\ref{A3VB1B3-F1}) into (\ref{EOM-V-TALE}), we can solve for the vector superfield $V$,
\begin{align}
\e^{- 2 V}
\ &= \ 
\frac{1}{|B_1|^2 + |B_3|^2} \Big\{
\frac{|B_1|^2 + |B_3|^2}{|B_3|^2} \frac{\F_1^2}{4}
- \sqrt{2} \, t_1
\Big\}
\, . \label{Dual1-e2V} 
\end{align}
Substituting the above result into the duality relation (\ref{A1Y1_A1-ALE}),
we find 
\begin{align}
|A_1|^2 \ &= \ 
\frac{\F_1^2}{4} \, \e^{-2 V}
\ = \ 
\frac{|A_3|^2 |B_3|^2}{|B_1|^2}
\nn \\
\ &= \ 
\frac{\beta_1 (Y_1 + \ol{Y}{}_1) - \sqrt{2} \, (A_3 B_3 C + \ol{A}{}_3 \ol{B}{}_3 \ol{C})}{2 (|B_1|^2 + |B_3|^2)} 
\Big\{
- \sqrt{2} \, t_1 
+ \sqrt{(\sqrt{2} \, t_1)^2 + \frac{4 |A_3|^2}{|B_1|^2} (|B_1|^2 + |B_3|^2)^2} 
\Big\}
\, . \label{A1Y1_A1-ALE-2}
\end{align}

Now we are ready to describe the supersymmetric NLSM without using the vector multiplet $\{ V, C \}$.
Substituting the above descriptions into the GLSM (\ref{DT-GLSMF_A1-ALE}) in the IR limit, we obtain 
\begin{align}
\Scr{L} \ &= \ 
- \int \d^4 \theta \, 
\beta_1 (Y_1 + \ol{Y}{}_1) 
\Big( \log |A_3|^2 - \log |B_1|^2 + \log |B_3|^2 \Big)
- \sqrt{2} \, t_1 \int \d^4 \theta \, \log{(|B_1|^2 + |B_3|^2)} 
\nn \\
\ & \ \ \ \ 
+ \int \d^4 \theta \, 
\sqrt{(\sqrt{2} \, t_1)^2 + \frac{4 |A_3|^2}{|B_1|^2} (|B_1|^2 + |B_3|^2)^2} 
\nn \\
\ & \ \ \ \ 
+ \sqrt{2} \, t_1 \int \d^4 \theta \, 
\log \left\{
- \sqrt{2} \, t_1 
+ \sqrt{(\sqrt{2} \, t_1)^2 + \frac{4 |A_3|^2}{|B_1|^2} (|B_1|^2 + |B_3|^2)^2} 
\right\}
\nn \\
\ & \ \ \ \ 
- \sqrt{2} \, t_2 F_{01}
\, . \label{Dual1C-SNLSM}
\end{align}
This is the NLSM dual to (\ref{GLSMF44-IR2}) in the superfield formalism.
We emphasize that this is the K\"{a}hler potential of the dualized system.
The K\"{a}hler potential will provide not only the target space geometry but also the target space B-field.
Since the K\"{a}hler potential (\ref{Dual1C-SNLSM}) and the duality relations (\ref{A1Y1_A1-ALE-2}) are too complicated to solve,
we do not expand it in terms of the component fields.
Instead, we will describe the GLSM (\ref{DT-GLSMF_A1-ALE}) in the language of the bosonic component fields, and analyze its low energy limit.

\subsubsection{Analysis by component fields}

We will now investigate the dual GLSM (\ref{DT-GLSMF_A1-ALE}) and its low energy limit in the language of the component fields.
In order to make the discussion clear, we do not describe the explicit forms of 
the duality relations and the dual Lagrangian in terms of the component fields in this subsection.
Instead they are given in
(\ref{DR-gen-1}) and (\ref{LDT-comp}) in appendix \ref{app:expand}.

Imposing the field equations for all the auxiliary fields,
we consider the supersymmetric Higgs branch.
Even under the duality transformation to the GLSM, 
we obtain the same constraints as (\ref{HB-ALE}), 
where the scalar fields $a_1$ is dual to $y_1$ 
under the duality relations (\ref{DR-gen-1}).
Furthermore, in analogy with section \ref{subsect:522}, 
we should impose the following equation as a part of the constraints on the supersymmetric Higgs branch,
\begin{align}
0 \ &= \ 
\big( a_1 b_1 A_{c=} + \ol{a}{}_1 \ol{b}{}_1 \ol{A}{}_{c=} \big)
\big( a_1 b_1 B_{c\+} + \ol{a}{}_1 \ol{b}{}_1 \ol{B}{}_{c\+} \big)
\, . \label{HB-TALE}
\end{align}
For convenience, we introduce the following reparametrization,
\begin{gather}
a_i \ \equiv \ \X_i \, \e^{\i \gamma_i}
\, , \ls
b_i \ \equiv \ \Y_i \, \e^{\i \lambda_i}
\, , \ls
y_1 \ \equiv \ \half (\U_1 + \i \V_1)
\, . \label{aby-gen}
\end{gather}
Then, the duality relations (\ref{DR-gen}) with the constraints (\ref{HB-ALE}) and (\ref{HB-TALE}) are described as
\bsubeq \label{constr-TALE}
\begin{align}
\beta_1 \, \U_1
\ &= \ 
\X_1^2 
- \sqrt{2} \, \X_1 \Y_1 \Big\{
\phi_c \, \e^{\i (\gamma_1 + \lambda_1)} 
+ \ol{\phi}{}_c \, \e^{- \i (\gamma_1 + \lambda_1)} 
\Big\}
\, , \label{a2HK-scalar3} \\
\beta_1 (\del_0 + \del_1) \V_1
\ &= \ 
\sqrt{2} \, \X_1 \Big\{
\sqrt{2} \, \X_1 
- \Y_1 \phi_c \, \e^{\i (\gamma_1 + \lambda_1)} 
- \Y_1 \ol{\phi}{}_c \, \e^{- \i (\gamma_1 + \lambda_1)} 
\Big\}
\, (\del_0 + \del_1) \gamma_1
\nn \\
\ & \ \ \ \ 
- \sqrt{2} \, \X_1 \Y_1 
\Big\{ \phi_c \, \e^{\i (\gamma_1 + \lambda_1)} 
+ \ol{\phi}{}_c \, \e^{- \i (\gamma_1 + \lambda_1)} 
\Big\}
\, (\del_0 + \del_1) \lambda_1 
\nn \\
\ & \ \ \ \ 
+ \i \sqrt{2} \, \Y_1 
\Big\{
\phi_c \, \e^{\i (\gamma_1 + \lambda_1)} 
- \ol{\phi}{}_c \, \e^{- \i (\gamma_1 + \lambda_1)} 
\Big\}
\, (\del_0 + \del_1) \X_1 
\nn \\
\ & \ \ \ \ 
+ \i \sqrt{2} \, \X_1 
\Big\{ \phi_c \, \e^{\i (\gamma_1 + \lambda_1)} 
- \ol{\phi}{}_c \, \e^{- \i (\gamma_1 + \lambda_1)} 
\Big\}
\, (\del_0 + \del_1) \Y_1 
\nn \\
\ & \ \ \ \ 
+ 2 \X_1^2 (A_0 + A_1)
- \sqrt{2} \, \X_1 \Y_1 B_{c\+} \, \e^{\i (\gamma_1 + \lambda_1)} 
- \sqrt{2} \, \X_1 \Y_1 \ol{B}{}_{c\+} \, \e^{- \i (\gamma_1 + \lambda_1)} 
\, , \\
- \beta_1 (\del_0 - \del_1) \V_1 
\ &= \ 
\sqrt{2} \, \X_1 
\Big\{
\sqrt{2} \, \X_1 
- \Y_1 \phi_c \, \e^{\i (\gamma_1 + \lambda_1)} 
- \Y_1 \ol{\phi}{}_c \, \e^{- \i (\gamma_1 + \lambda_1)} 
\Big\}
\, (\del_0 - \del_1) \gamma_1
\nn \\
\ & \ \ \ \ 
- \sqrt{2} \, \X_1 \Y_1 
\Big\{ 
\phi_c \, \e^{\i (\gamma_1 + \lambda_1)} 
+ \ol{\phi}{}_c \, \e^{- \i (\gamma_1 + \lambda_1)} 
\Big\}
\, (\del_0 - \del_1) \lambda_1
\nn \\
\ & \ \ \ \ 
+ \i \sqrt{2} \, \Y_1 
\Big\{
\phi_c \, \e^{\i (\gamma_1 + \lambda_1)} 
- \ol{\phi}{}_c \, \e^{- \i (\gamma_1 + \lambda_1)} 
\Big\}
\, (\del_0 - \del_1) \X_1 
\nn \\
\ & \ \ \ \ 
+ \i \sqrt{2} \, \X_1 
\Big\{
\phi_c \, \e^{\i (\gamma_1 + \lambda_1)} 
- \ol{\phi}{}_c \, \e^{- \i (\gamma_1 + \lambda_1)} 
\Big\}
\, (\del_0 - \del_1) \Y_1 
\nn \\
\ & \ \ \ \ 
+ 2 \X_1^2 (A_0 - A_1)
- \sqrt{2} \, \X_1 \Y_1 A_{c=} \, \e^{\i (\gamma_1 + \lambda_1)} 
- \sqrt{2} \, \X_1 \Y_1 \ol{A}{}_{c=} \, \e^{- \i (\gamma_1 + \lambda_1)} 
\, , \\
0 \ &= \ 
(\X_1^2 + \X_3^2) - (\Y_1^2 + \Y_3^2) - \frac{\a^2}{2}
\, , \\
0 \ &= \ 
\X_1 \Y_1 \, \e^{\i (\gamma_1 + \lambda_1)}
+ \X_3 \Y_3 \, \e^{\i (\gamma_3 + \lambda_3)}
\, , \\
0 \ &= \ 
\big( A_{c=} \, \e^{\i (\gamma_1 + \lambda_1)} 
+ \ol{A}{}_{c=} \, \e^{- \i (\gamma_1 + \lambda_1)} \big)
\big( B_{c\+} \, \e^{\i (\gamma_1 + \lambda_1)} 
+ \ol{B}{}_{c\+} \, \e^{- \i (\gamma_1 + \lambda_1)} \big)
\, .
\end{align}
\esubeq
Although the existence of the fields $\{ \phi_c, A_{c=}, B_{c\+} \}$ implies that the contribution of $C$ to the system is still remaining,
the above expression is too hard to analyze.
So we consider the restriction to the Higgs branch,
\begin{align}
\phi_c \ = \ 0
\, . \label{special-Higgs-TALE}
\end{align}
This is a strong constraint. 
However, we will find the correct NLSM in the end.
The equations (\ref{constr-TALE}) are drastically reduced to
\bsubeq \label{A1Y1_A1-ALE-comp}
\begin{align}
\beta_1 \U_1
\ &= \ 
\X_1^2
\, , \\
\beta_1 (\del_0 + \del_1) \V_1
\ &= \ 
2 \X_1^2 \big\{ (\del_0 + \del_1) \gamma_1 + (A_0 + A_1) \big\}
\nn \\
\ & \ \ \ \ 
- \sqrt{2} \, \X_1 \Y_1 \big\{
B_{c\+} \, \e^{\i (\gamma_1 + \lambda_1)} 
+ \ol{B}{}_{c\+} \, \e^{- \i (\gamma_1 + \lambda_1)} 
\big\}
\, , \\
- \beta_1 (\del_0 - \del_1) \V_1 
\ &= \ 
2 \X_1^2 \big\{ (\del_0 - \del_1) \gamma_1 + (A_0 - A_1) \big\}
\nn \\
\ & \ \ \ \ 
- \sqrt{2} \, \X_1 \Y_1 \big\{
A_{c=} \, \e^{\i (\gamma_1 + \lambda_1)} 
+ \ol{A}{}_{c=} \, \e^{- \i (\gamma_1 + \lambda_1)} 
\big\}
\, , 
\end{align}
and 
\begin{align}
0 \ &= \ 
(\X_1^2 + \X_3^2) - (\Y_1^2 + \Y_3^2) - \frac{\a^2}{2}
\, , \\
0 \ &= \ 
\X_1 \Y_1 \, \e^{\i (\gamma_1 + \lambda_1)}
+ \X_3 \Y_3 \, \e^{\i (\gamma_3 + \lambda_3)}
\, , \\
0 \ &= \ 
\big( A_{c=} \, \e^{\i (\gamma_1 + \lambda_1)} 
+ \ol{A}{}_{c=} \, \e^{- \i (\gamma_1 + \lambda_1)} \big)
\big( B_{c\+} \, \e^{\i (\gamma_1 + \lambda_1)} 
+ \ol{B}{}_{c\+} \, \e^{- \i (\gamma_1 + \lambda_1)} \big)
\nn \\
\ &= \ 
\Big\{
(\del_0 + \del_1) \gamma_1
- \frac{\beta_1}{2 \X_1^2} (\del_0 + \del_1) \V_1
+ (A_0 + A_1)
\Big\}
\Big\{
(\del_0 - \del_1) \gamma_1
+ \frac{\beta_1}{2 \X_1^2} (\del_0 - \del_1) \V_1
+ (A_0 - A_1)
\Big\}
\, . \label{V2g-TALE}
\end{align}
\esubeq
Plugging these equations into the GLSM (\ref{DT-GLSMF_A1-ALE}), 
we obtain the simplified form,
\begin{align}
\Scr{L}
\ &= \ 
\frac{1}{2 e^2} (F_{01})^2 
+ \frac{1}{2 \wt{e}^2} (\wt{F}_{01})^2 
\nn \\
\ & \ \ \ \ 
- (\X_1^2 + \X_3^2 + \Y_1^2 + \Y_3^2) A_m A^m
- 2 \Big\{ 
\X_1^2 (\del^m \gamma_1) + \X_3^2 (\del^m \gamma_3) 
- \Y_1^2 (\del^m \lambda_1) - \Y_3^2 (\del^m \lambda_3) 
\Big\} A_m 
\nn \\
\ & \ \ \ \ 
- (\del_m \X_1)^2
- (\del_m \X_3)^2
- (\del_m \Y_1)^2
- (\del_m \Y_3)^2
\nn \\
\ & \ \ \ \ 
- \X_1^2 (\del_m \gamma_1)^2
- \X_3^2 (\del_m \gamma_3)^2
- \Y_1^2 (\del_m \lambda_1)^2
- \Y_3^2 (\del_m \lambda_3)^2
- \ve^{mn} (\del_m \gamma_1) \, \del_n (\beta_1 \V_1)
\nn \\
\ & \ \ \ \ 
- \sqrt{2} \, t_2 F_{01}
+ \text{(fermionic fields)}
\, . \label{Dual1C-GLSMF-2}
\end{align}
Owing to the existence of the auxiliary fields $\{ A_{c=}, B_{c\+} \}$,
the second derivative of the original scalar field $\gamma_1$ is revived, whilst 
the second derivative of the dual scalar field $\V_1$ disappears.
The same phenomenon has been discussed in (\ref{add-HB-522}) in section \ref{subsect:522}.
We have to keep in mind that 
$\gamma_1$ is no longer the dynamical field 
even though it has a second derivative term in (\ref{Dual1C-GLSMF-2}).
Instead, the field $\V_1$ is the new dynamical field via the duality relations (\ref{A1Y1_A1-ALE-comp}), or more concretely via the equation (\ref{V2g-TALE}).

Since we would like to obtain the string worldsheet sigma model in the IR limit of the GLSM, 
we take the IR limit $e, \wt{e} \to \infty$.
Then the gauge field $\wt{A}_m$ is decoupled from the system and $A_m$ becomes an auxiliary field.
We can solve the field equation for $A_m$,
\begin{align}
A_m \ &= \ 
- \frac{2}{\rho^2}
\Big\{
\X_1^2 (\del_m \gamma_1) + \X_3^2 (\del_m \gamma_3) 
- \Y_1^2 (\del_m \lambda_1) - \Y_3^2 (\del_m \lambda_3) 
\Big\}
\, . \label{EOM-A-AY1-Cver}
\end{align}
Plugging this into the Lagrangian (\ref{Dual1C-GLSMF-2}) 
and introducing the parametrization (\ref{A13B13-GLSM}),
we obtain 
\begin{align}
\Scr{L}
\ &= \ 
- \half \Big( 1 - \frac{\a^4}{\rho^4} \Big)^{-1} (\del_m \rho)^2
- \frac{\rho^2}{8} (\del_m \vartheta)^2
- \frac{\rho^4 - \a^4}{8 \rho^2} (\del_m \psi)^2 
\nn \\
\ & \ \ \ \ 
- \frac{\rho^4 - \a^4 \cos^2 \vartheta}{8 \rho^2} (\del_m \varphi)^2 
- \frac{(\rho^4 - \a^4) \cos \vartheta}{4 \rho^2} (\del_m \psi) (\del^m \varphi)
- \half \ve^{mn} \big\{ \wt{a} \, \del_m \psi + \del_m \varphi \big\} \, \del_n (\beta_1 \V_1) 
\nn \\
\ & \ \ \ \ 
- \sqrt{2} \, t_2 F_{01}
+ \text{(fermionic fields)}
\, . \label{Dual1C-GLSMF-2-3}
\end{align}
Finally, we should integrate out the original non-dynamical scalar field $\varphi$ 
using the equation of motion,
\begin{align}
\del_m \varphi
\ &= \ 
- \frac{2 \rho^2}{\rho^4 - \a^4 \cos^2 \vartheta} \Big\{
\ve_{mn} \, \del^n (\beta_1 \V_1) 
+ \frac{(\rho^4 - \a^4) \cos \vartheta}{2 \rho^2} (\del_m \psi) 
\Big\}
+ X_m
\, ,
\end{align}
where $X_m$ is an integration constant vector.
This should vanish because of the Lorentz invariance in two-dimensional space.
Substituting this into (\ref{Dual1C-GLSMF-2-3}), 
we obtain the final form of the NLSM,
\begin{align}
\Scr{L}
\ &= \ 
- \frac{\A^{-1}}{2} (\del_m \rho)^2
- \frac{\rho^2}{8} (\del_m \vartheta)^2
- \frac{\rho^2}{4} \frac{\A \sin^2 \vartheta}{\A \cos^2 \vartheta + \sin^2 \vartheta}
(\del_m \psi)^2
- \frac{\beta_1^2}{2 \rho^2} \frac{1}{\A \cos^2 \vartheta + \sin^2 \vartheta}
(\del_m \V_1)^2
\nn \\
\ & \ \ \ \ 
- \frac{\beta_1}{2} 
\frac{\A \cos \vartheta}{\A \cos^2 \vartheta + \sin^2 \vartheta} 
\, \ve^{mn} \, (\del_m \V_1) (\del_n \psi) 
+ \frac{\wt{a} \beta_1}{2} \, \ve^{mn} \, (\del_m \V_1) (\del_n \psi) 
\nn \\
\ & \ \ \ \ 
- \sqrt{2} \, t_2 F_{01}
+ \text{(fermionic fields)}
\, ,
\end{align}
where $\A = 1 - \a^4/\rho^4$.
We can remove the second term in the second line of the right-hand side 
because this is a total derivative term.
It turns out that the target space of this NLSM corresponds to 
the T-duality configuration of the $A_1$-type ALE space (\ref{Tdual-EH})
via the Buscher rule
if we set $\beta_1 = - 2$ and $\V_1 = \varphi'$.
We conclude that the duality transformation procedure discussed in section \ref{subsect:DT-CCS-DF} is completely correct 
for describing the T-duality transformation in the superfield formalism.

We remark that the configuration of the target space of the NLSM represents 
two parallel NS5-branes with non-vanishing B-field (for the detailed discussion, see appendix \ref{app:EH}).
On the other hand, we have also obtained the equivalent configuration by the NLSM (\ref{GLSM-Hm-b3}) with $k = 2$ under the small $R_a$ limit in (\ref{H-omega-Hm}).
This implies that the physics describing the GLSM (\ref{DT-GLSMF_A1-ALE}) corresponds to that of the GLSM (\ref{GLSM-HM}).

We conclude that the duality transformation rule in the presence of the F-term discussed in section \ref{subsect:DT-CCS-DF} is consistent with the T-duality transformation rule,
and again confirmed that 
the existence of the F-term and its contribution to the duality transformation rule in the GLSMs are required
to reproduce the correct T-duality transformations.

\section{Summary and discussions}
\label{sect:summary}


In this paper we studied the duality transformation rules of two-dimensional chiral superfields in the D-term and the F-term.
The technique can be briefly summarized as follows.
If the chiral superfield $\Psi$ which we dualize is coupled to another chiral superfield $\Phi$ in F-term,
we first express $\Phi$ in terms of an unconstrained complex superfield $C$.
Then we convert the F-term to D-terms.
Once all the interaction terms with kinetic terms only appear
in the D-terms, we perform the conventional duality transformation discussed in \cite{Rocek:1991ps} or \cite{Hori:2000kt}.
In this work, in particular, 
we demonstrated the duality transformation of the neutral chiral superfield in the D-term and the F-term, and the duality transformation of the charged chiral superfield in D-term and F-term.
In order to justify the duality transformations to the T-duality transformations in the superfield formalism,
we explained two typical examples:
One is the $\N=(4,4)$ GLSM for five-branes, 
and the other is the $\N=(4,4)$ GLSM for the $A_1$-type ALE space.
The former has been discussed in \cite{Tong:2002rq} and \cite{Kimura:2013fda},
while the latter is a novel application.
Because we successfully found that the dual NLSM represents the string sigma model on the T-dualized configuration, 
we concluded that the duality transformation procedures are consistent.
Since the procedures are quite general,
we believe that they could be applied to many topics in string theory.

In this work we obtained two important results.
The first one is the explicit form of the K\"{a}hler potential in the dual system.
By virtue of the new duality transformation,
we obtained the explicit expression of the K\"{a}hler potential of the configuration of two parallel NS5-branes 
with non-vanishing B-field in ten-dimensional supergravity.
The second important result is that 
we obtained two formulations for two parallel NS5-branes: 
one is the $\N=(4,4)$ GLSM (\ref{GLSM-HM}) with $k=2$,
and the other is the $\N=(4,4)$ GLSM (\ref{DT-GLSMF_A1-ALE}).
They are equivalent to each other in the IR limit.
Analogously, we also obtained two equivalent formulations for two parallel KK-monopoles:
the GLSM (\ref{GLSM-KKM}) with $k=2$ and the GLSM (\ref{GLSMF_A1-ALE}).


In this work we only investigated the Higgs branch of the GLSM.
In particular, in the analysis of the dual GLSM for the $A_1$-type ALE space
we further restricted the system with $\phi_c = 0$ (\ref{special-Higgs-TALE}).
This is a strong condition, even though we could find the T-dualized configuration of the ALE space.
In order to understand the whole structure of the GLSM and its T-dualized model,
we have to remove this condition within the Higgs branch. 
We should also analyze the Coulomb branch, 
which could connect two distinct features of the NLSMs in the IR limit, as in the case of GLSMs for Calabi-Yau varieties \cite{Witten:1993yc}.

It is interesting to apply our techniques in this work to other topics in string theory.
For instance, we could apply our techniques to the gauge theoretical construction of NLSMs on non-compact Calabi-Yau manifolds discussed in 
\cite{Higashijima:2001yn, Higashijima:2001fp, Higashijima:2002px, Kimura:2004ds}.
More attractively, 
double field theory is a candidate 
because this theory involves various string theories connected to each other via T-duality transformations in a beautiful way (see, for instance, \cite{Hull:2009mi, Hohm:2010jy, Hohm:2010pp, Lust:2010iy, Jensen:2011jna, Hohm:2011dv, Kikuchi:2012za, Hassler:2014sba, Rennecke:2014sca} and recent reviews \cite{Zwiebach:2011rg, Aldazabal:2013sca, Berman:2013eva}).
It is also interesting to apply our techniques to string worldsheet theory with a boundary.
It is well known that the worldsheet boundary plays a crucial role in describing D-branes.
D-branes are sources of Ramond-Ramond potentials 
which are difficult to describe as operators in the string worldsheet.
However, D-branes are non-trivially transformed under the T-duality transformations. 
There are several works investigating
the boundary of string worldsheet sigma models
(for instance, see \cite{Hori:2000ck, Albertsson:2008gq, Albertsson:2011ux}).

Finally, we should remark on the validity of our techniques.
As long as we only consider
abelian T-duality transformations, i.e.
the T-duality transformations along abelian isometries,
the techniques discussed in section \ref{sect:DTR} are completely valid.
However, they are no longer applicable to investigate ``non-abelian'' T-duality transformations.
It seems interesting to import previous work on
the non-abelian T-duality procedure \cite{delaOssa:1992vc, Giveon:1993ai} to our techniques.

\section*{Acknowledgements}

The authors would like to thank 
Shun'ya Mizoguchi 
and
Shin Sasaki
for helpful discussions.
They also thank Christopher Locke for his careful reading of this manuscript.
TK is grateful to the Yukawa Institute for Theoretical Physics at
Kyoto University. Discussions during the YITP molecule-type workshop on
``Exotic Structures of Spacetime'' (YITP-T-13-07) were useful to
complete this work. 
The work of TK is supported in part by the Iwanami-Fujukai Foundation.

\begin{appendix}
\section*{Appendix}

\section{Conventions}
\label{app:conventions}

In this appendix we outline the conventions for two-dimensional supersymmetric field theories.

\subsection{String worldsheet sigma model}

Here we will define the normalization of the string worldsheet sigma model.
This is important for discussing the T-duality transformations of
the configuration of the target space.
The action and the Lagrangian is described in the standard form,
\bsubeq \label{string-NLSM-bosons}
\begin{gather}
S \ = \ 
\frac{1}{2 \pi \alpha'} \int \d^2 \sigma \, \Scr{L}
\, , \\
\Scr{L}
\ = \ 
- \half G_{IJ} \, g^{mn} \, \del_m X^I \del_n X^J
+ \half B_{IJ} \, \ve^{mn} \, \del_m X^I \del_n X^J
\, . 
\end{gather}
\esubeq
Here $G_{IJ}$ and $B_{IJ}$ denote the target space metric and the B-field
respectively,
and they should follow the equations of motion of supergravity theories.
The target space dilaton does not appear in this sigma model 
if the worldsheet metric $g_{mn}$ is flat.
We set the normalization of the metric $g_{mn}$ and the Levi-Civita invariant tensor $\ve_{mn}$ on the flat space to $g_{mn} = \text{diag.}(-,+)$ and $\ve^{01} = +1 = \ve^{10}$ respectively.
For convenience, we set $2 \pi \alpha' = 1$.

\subsection{$\N=(2,2)$ supersymmetry in two dimensions}
\label{app:N=22SUSY}

In section \ref{sect:EB} and section \ref{sect:TA-ALE} 
we discussed two-dimensional $\N=(4,4)$ GLSMs and their IR limit.
It is quite useful to set up $\N=(2,2)$ supersymmetry
because the two-dimensional $\N=(2,2)$ supersymmetry can be realized as the dimensional reduction from four-dimensional $\N=1$ supersymmetry
which is quite common in quantum field theory.

We begin with the definition of the supercovariant derivatives in 
superspace expanded by 
supercoordinates $\{ x^m , \theta^{\alpha} , \ol{\theta}{}^{\alpha} \}$,
\begin{alignat}{2}
D_{\pm} 
\ &= \
\frac{\del}{\del \theta^{\pm}}
- \i \ol{\theta}{}^{\pm} 
(\del_0 \pm \del_1)
\, , &\ls
\ol{D}{}_{\pm}
\ &= \ 
- \frac{\del}{\del \ol{\theta}{}^{\pm}}
+ \i \theta^{\pm} 
(\del_0 \pm \del_1)
\, . 
\end{alignat}
Here the variables $\theta^{\alpha}$ are the Grassmann-odd coordinates in superspace.
we adopted the spinor indices $\alpha = \pm$,
which is suitable for light-cone coordinates in two-dimensional spacetime.

In order to construct the conventional Lagrangian in the superspace formalism,
we have to define the integral measures of the Grassmann-odd coordinates.
In this work the normalization of the integral measures are given as
\begin{align}
\d^2 \theta 
\ &= \ 
- \half \d \theta^+ \d \theta^-
\, , \ls
\d^2 \wt{\theta}
\ = \ 
- \half \d \theta^+ \d \ol{\theta}{}^-
\, , \ls
\d^4 \theta 
\ = \ 
- \frac{1}{4} \d \theta^+ \d \theta^- \d \ol{\theta}{}^+ \d \ol{\theta}{}^-
\, .
\end{align}

We introduce various irreducible superfields in two-dimensional theories.
We define chiral superfields and their expansions in terms of component fields
which play a central role in the duality transformations in this 
work,
\bsubeq
\begin{align}
A_i \ &= \ 
a_i 
+ \i \sqrt{2} \, \theta^+ \psi_{i,+} 
+ \i \sqrt{2} \, \theta^- \psi_{i,-} 
+ 2 \i \, \theta^+ \theta^- F_i  
+ \text{(derivative terms)}
\, , \\
B_i \ &= \ 
b_i 
+ \i \sqrt{2} \, \theta^+ \wt{\psi}_{i,+} 
+ \i \sqrt{2} \, \theta^- \wt{\psi}_{i,-} 
+ 2 \i \, \theta^+ \theta^- \wt{F}_i  
+ \text{(derivative terms)}
\, .
\end{align}
\esubeq
The derivative terms are determined by the definition $\ol{D}{}_{\pm} A_i = 0 = \ol{D}{}_{\pm} B_i$.
We also introduce a twisted chiral superfield $Y_i$ and its expansion in terms of 
component fields, which is another main ingredient in the duality 
transformations,
\begin{align}
Y_i \ &= \ 
y_i
+ \i \sqrt{2} \, \theta^+ \ol{\chi}{}_{i,+} 
+ \i \sqrt{2} \, \ol{\theta}{}^- \chi_{i,-} 
+ 2 \i \, \theta^+ \ol{\theta}{}^- G_i 
+ \text{(derivative terms)}
\, .
\end{align}
The derivative terms are determined by the definition $\ol{D}{}_{+} Y_i = 0 = D_- Y_i$.

We can construct $\N=(4,4)$ supersymmetric multiplets such as hypermultiplets
and vector multiplets 
as doublets of $\N=(2,2)$ superfields under the $SU(2)_R$ symmetry.
For instance, a set of $\N=(2,2)$ chiral superfields $\{ A_i, B_i \}$ can be utilized as an $SU(2)_R$ doublet.
In this example, the $SU(2)_R$ rotation acting on the component fields is expressed in the following way,
\begin{align}
(a_i, \ol{b}{}_i) \ &\to \ 
(\ol{b}{}_i, - a_i)
\, , \ls
(\psi_{i,\pm}, \ol{\wt{\psi}}{}_{i,\pm})
\ \to \ 
(\psi_{i,\pm}, \ol{\wt{\psi}}{}_{i,\pm})
\, .
\end{align}

We construct an $\N=(4,4)$ abelian vector multiplet 
in terms of an $\N=(2,2)$ vector superfield $V$ and a neutral chiral superfield $\Phi$.
The pair is also a doublet of the $SU(2)_R$ symmetry in $\N=(4,4)$ supersymmetry.
There are many redundant degrees of freedom in this multiplet.
Thus we take the Wess-Zumino gauge to reduce them.
The explicit expansions of the superfields under this gauge are given as
\bsubeq
\begin{align}
V \ &= \ 
\theta^+ \ol{\theta}{}^+ (A_0 + A_1)
+ \theta^- \ol{\theta}{}^- (A_0 - A_1)
- \sqrt{2} \, \theta^- \ol{\theta}{}^+ \sigma 
- \sqrt{2} \, \theta^+ \ol{\theta}{}^- \ol{\sigma}
\nn \\
\ & \ \ \ \ 
- 2 \i \, \theta^+ \theta^- 
\big( \ol{\theta}{}^+ \ol{\lambda}{}_+ + \ol{\theta}{}^- \ol{\lambda}{}_- \big)
+ 2 \i \, \ol{\theta}{}^+ \ol{\theta}{}^- 
\big( \theta^+ \lambda_+ + \theta^- \lambda_- \big)
- 2 \theta^+ \theta^- \ol{\theta}{}^+ \ol{\theta}{}^- D_V 
\, , \\
\Sigma \ &= \ 
\frac{1}{\sqrt{2}} \ol{D}{}_+ D_- V
\nn \\
\ &= \ 
\sigma 
- \i \sqrt{2} \, \theta{}^+ \ol{\lambda}{}_+ 
- \i \sqrt{2} \, \ol{\theta}{}^- \lambda_- 
+ \sqrt{2} \, \theta{}^+ \ol{\theta}{}^- (D_V - \i F_{01}) 
+ \text{(derivative terms)}
\, , \\
\Phi \ &= \ 
\phi
+ \i \sqrt{2} \, \theta^+ \wt{\lambda}_{+} 
+ \i \sqrt{2} \, \theta^- \wt{\lambda}_{-}
+ 2 \i \, \theta^+ \theta^- D_{\Phi}
+ \text{(derivative terms)}
\, .
\end{align}
\esubeq
The $SU(2)_R$ rotation acting on the component fields of the $\N=(4,4)$ vector multiplet is given as
\begin{align}
(\sigma, \phi) \ &\to \ 
(\sigma, \phi)
\, , \ls
(\lambda_{\pm}, \wt{\lambda}_{\pm})
\ \to \ 
(\wt{\lambda}_{\pm}, - \lambda_{\pm})
\, . 
\end{align}
In order to preserve this rotation, 
we determine the coefficient $\wt{\alpha} = \alpha$ in the F-term of (\ref{LDF-CCS}).

In the main part of this paper
we rewrite the neutral chiral superfield $\Phi$ in terms of the unconstrained complex superfield $C$ in such a way that 
$\Phi = \ol{D}{}_+ \ol{D}{}_- C$.
It is worth describing the explicit expansion of this superfield in terms of the component fields,
\begin{align}
C
\ &= \ 
\phi_{c} 
+ \i \sqrt{2} \, \theta^+ \psi_{c+} 
+ \i \sqrt{2} \, \theta^- \psi_{c-} 
+ \i \sqrt{2} \, \ol{\theta}{}^+ \chi_{c+} 
+ \i \sqrt{2} \, \ol{\theta}{}^- \chi_{c-}
\nn \\
\ & \ \ \ \ 
+ 2 \i \, \theta^+ \theta^- F_{c} 
+ 2 \i \, \ol{\theta}{}^+ \ol{\theta}{}^- M_{c}
+ 2 \i \, \theta^+ \ol{\theta}{}^- G_{c} 
+ 2 \i \, \ol{\theta}{}^+ \theta^- N_{c}
+ \theta^- \ol{\theta}{}^- A_{c=}
+ \theta^+ \ol{\theta}{}^+ B_{c\+}
\nn \\
\ & \ \ \ \ 
- 2 \i \, \theta^+ \theta^- \ol{\theta}{}^+ \zeta_{c+}
- 2 \i \, \theta^+ \theta^- \ol{\theta}{}^- \zeta_{c-}
+ 2 \i \, \ol{\theta}{}^+ \ol{\theta}{}^- \theta^+ \lambda_{c+}
+ 2 \i \, \ol{\theta}{}^+ \ol{\theta}{}^- \theta^- \lambda_{c-}
\nn \\
\ & \ \ \ \ 
- 2 \theta^+ \theta^- \ol{\theta}{}^+ \ol{\theta}{}^- D_{c}
\, .
\end{align}
The relation among the component fields of $\Phi$ and $C$ is given as\footnote{Notice that the relation among $\{ D_{\Phi}, D_c, A_{c=}, B_{c\+}, \phi_c \}$ in \cite{Kimura:2013fda, Kimura:2013zva, Kimura:2013khz} has an error in the coefficient of $A_{c=}$.}
\bsubeq \label{Phi2C}
\begin{align}
\phi
\ &= \ 
- 2 \i \, M_{c} 
\, , \\
D_{\Phi}
\ &= \
- \i \, D_{c}
+ \frac{1}{2} (\del_0 + \del_1) A_{c=}
+ \frac{1}{2} (\del_0 - \del_1) B_{c\+}
+ \frac{\i}{2} (\del_0^2 - \del_1^2) \phi_{c}
\, , \\
\wt{\lambda}_{\pm}
\ &= \ 
- \sqrt{2} \, \lambda_{c\pm}
\mp \i (\del_0 \pm \del_1) \chi_{c\mp}
\, ,
\end{align}
\esubeq
while there are no relations for the component fields 
$\{ F_{c}, G_{c}, N_{c}, \psi_{c\pm}, \zeta_{c\pm} \}$.

\section{Explicit expansions in terms of bosonic component fields}
\label{app:expand}

In this appendix we discuss the explicit expressions of the duality relation and the dual Lagrangian in terms of the bosonic component fields.
Here we neglect all the fermionic component fields.
By using the expansion rules of the superfields, 
the explicit forms of the duality relation
(\ref{AY-dual-CCS}) or (\ref{A1Y1_A1-ALE})
among the bosonic component fields are given as follows,
\bsubeq \label{DR-gen}
\begin{align}
|a_i|^2 
\ &= \ 
\beta_i (y_i + \ol{y}{}_i)
+ \sqrt{2} \, \H_{i,+}
\, , \label{DR-gen-1} \\
F_i   
\ &= \ 
\frac{2 a_i}{\K_i - \H_{i,-}}
\Big\{ 
a_i \phi_c \wt{F}_i  
+ a_i b_i F_{c} 
+ \ol{a}{}_i \ol{b}{}_i \ol{M}{}_{c} 
\Big\}
\, , \\
\ol{F}_i 
\ &= \ 
\frac{2 \ol{a}{}_i}{\K_i + \H_{i,-}}
\Big\{ 
\ol{a}{}_i \ol{\phi}{}_c \ol{\wt{F}}_i 
+ \ol{a}{}_i \ol{b}{}_i \ol{F}{}_{c} 
+ a_i b_i M_{c}
\Big\}
\, , \\
\ol{\sigma}
\ &= \ 
- \frac{\i}{\sqrt{2} \, |a_i|^2}
\Big\{
  \beta_i G_i 
+ \sqrt{2} \, a_i b_i G_{c} 
+ \sqrt{2} \, \ol{a}{}_i \ol{b}{}_i \ol{N}{}_{c} 
\Big\}
\, , \\
\sigma 
\ &= \ 
\frac{\i}{\sqrt{2} \, |a_i|^2}
\Big\{ 
  \beta_i \ol{G}{}_i
+ \sqrt{2} \, \ol{a}{}_i \ol{b}{}_i \ol{G}{}_{c}
+ \sqrt{2} \, a_i b_i N_{c}
\Big\}
\, , \\
\beta_i (\del_0 + \del_1) (y_i - \ol{y}{}_i)
\ &= \ 
- \frac{\K_i + \H_{i,-}}{\sqrt{2} \, |a_i|^2} \, a_i (\del_0 + \del_1) \ol{a}_i
+ \frac{\K_i - \H_{i,-}}{\sqrt{2} \, |a_i|^2} \, \ol{a}{}_i (\del_0 + \del_1) a_i
\nn \\
\ & \ \ \ \ 
- \sqrt{2} \, a_i \phi_c \, (\del_0 + \del_1) b_i
+ \sqrt{2} \, \ol{a}{}_i \ol{\phi}{}_c \, (\del_0 + \del_1) \ol{b}_i
+ 2 \i \, |a_i|^2 (A_0 + A_1)
\nn \\
\ & \ \ \ \ 
- \i \sqrt{2} \, a_i b_i B_{c\+}
- \i \sqrt{2} \, \ol{a}{}_i \ol{b}{}_i \ol{B}{}_{c\+}
\, , \\
- \beta_i (\del_0 - \del_1) (y_i - \ol{y}{}_i)
\ &= \ 
- \frac{\K_i + \H_{i,-}}{\sqrt{2} \, |a_i|^2} \, a_i (\del_0 - \del_1) \ol{a}_i
+ \frac{\K_i - \H_{i,-}}{\sqrt{2} \, |a_i|^2} \, \ol{a}{}_i (\del_0 - \del_1) a_i
\nn \\
\ & \ \ \ \ 
- \sqrt{2} \, a_i \phi_c \, (\del_0 - \del_1) b_i
+ \sqrt{2} \, \ol{a}{}_i \ol{\phi}{}_c \, (\del_0 - \del_1) \ol{b}_i
+ 2 \i \, |a_i|^2 (A_0 - A_1)
\nn \\
\ & \ \ \ \ 
- \i \sqrt{2} \, a_i b_i A_{c=}
- \i \sqrt{2} \, \ol{a}{}_i \ol{b}{}_i \ol{A}{}_{c=}
\, .
\end{align}
\esubeq
Here we have introduced the following functions,
\begin{align}
\H_{i,\pm}
\ &\equiv \ 
a_i b_i \phi_c \pm \ol{a}{}_i \ol{b}{}_i \ol{\phi}{}_c
\, , \ls
\K_i \ \equiv \ 
\sqrt{\H_{i,+}^2 + 2 \beta_i (y_i + \ol{y}{}_i) |a_i|^2}
\, . \label{H+K-generic-Ldual}
\end{align}
The key point is that the real part of the complex scalar field $y_i$ is related to other fields without any derivatives, 
whilst the derivative of the imaginary part of $y_i$ is related to 
(the derivatives of) other fields and the abelian vector field $A_m$.
Indeed, the derivatives play a crucial role in the duality transformation procedure.
We also note that the scalar field $\phi_c$ belonging to the unconstrained complex superfield $C$ appears in many terms in the duality relations.

We also discuss the expansion of the dual Lagrangian
given by
the first term in the right-hand side of (\ref{DT-LDF-CCS-1}), or 
the third line in the right-hand side of (\ref{DT-GLSMF_A1-ALE}),
\begin{align}
\Scr{L}_{\text{dual}}
\ &\equiv \ 
\int \d^4 \theta \, \Big\{
- 2 \beta_i (Y_i + \ol{Y}{}_i) \log \F_i
+ \half \F_i \T_i
\Big\}
\, . \label{LDT-part}
\end{align}
Because the explicit forms of $\F_i$ and $\T_i$ defined in (\ref{DT-LDF-CCS}) are complicated even in the superfield formalism,
we expand (\ref{LDT-part}) in terms of the component fields under the duality relations (\ref{DR-gen}),
\begin{align}
\Scr{L}_{\text{dual}}
\ &= \ 
\frac{\beta_i}{4} \log |a_i|^2 
\Big\{
\del_m \del^m (y_i + \ol{y}{}_i)
\Big\}
+ \frac{\beta_i^2}{2 \sqrt{2} \, \K_i} \, 
\big[ \del_m (y_i - \ol{y}{}_i) \big]^2
\nn \\
\ & \ \ \ \ 
- \frac{(\H_{i,-}^2 + \H_{i,+} \K_i)}{2 \sqrt{2} \, |a_i|^2 \K_i} \, 
|\del_m a_i|^2
- \frac{\H_{i,-}}{2 \sqrt{2} \, |a_i|^2} \, 
\Big\{ 
  \ol{a}{}_i \, (\del_m \del^m a_i) 
- a_i \, (\del_m \del^m \ol{a}_i) 
\Big\}
\nn \\
\ & \ \ \ \ 
- \frac{2 \ol{a}{}_i \ol{b}{}_i \ol{\phi}{}_c \K_i - (\K_i - \H_{i,-}) \H_{i,-}}{4 \sqrt{2} \, (a_i)^2 \K_i} \, 
(\del_m a_i)^2
- \frac{2 a_i b_i \phi_c \K_i + (\K_i + \H_{i,-}) \H_{i,-}}{4 \sqrt{2} \, (\ol{a}{}_i)^2 \K_i} \, 
(\del_m \ol{a}{}_i)^2
\nn \\
\ & \ \ \ \ 
+ \frac{\sqrt{2} \, |a_i|^2 |\phi_c|^2}{\K_i} \, |\del_m b_i|^2
- \frac{a_i \phi_c}{\sqrt{2}} \, 
(\del_m \del^m b_i)
- \frac{\ol{a}{}_i \ol{\phi}{}_c}{\sqrt{2}} \, 
(\del_m \del^m \ol{b}_i) 
- \frac{(a_i \phi_c)^2}{\sqrt{2} \, \K_i} \, (\del_m b_i)^2
- \frac{(\ol{a}{}_i \ol{\phi}{}_c)^2}{\sqrt{2} \, \K_i} \, (\del_m \ol{b}{}_i)^2 
\nn \\
\ & \ \ \ \ 
- \Big\{ 
\frac{\phi_c (\K_i + \H_{i,-})}{\sqrt{2} \, \ol{a}{}_i \K_i} \, (\del^m b_i) 
+ \frac{\ol{\phi}{}_c (\K_i - \H_{i,-})}{\sqrt{2} \, a_i \K_i} \, (\del^m \ol{b}{}_i) 
\Big\}
\Big( a_i \del_m \ol{a}{}_i + \ol{a}{}_i \del_m a_i \Big)
\nn \\
\ & \ \ \ \ 
+ \frac{\H_{i,-} \beta_i}{2 |a_i|^2 \K_i} \, 
\ve^{mn} \, \del_n (y_i - \ol{y}{}_i) 
\Big( a_i \del_m \ol{a}{}_i + \ol{a}{}_i \del_m a_i \Big)
+ \frac{\beta_i}{\K_i} \, 
\ve^{mn} \, \del_n (y_i - \ol{y}{}_i) 
\Big( a_i \phi_c \, \del_m b_i - \ol{a}{}_i \ol{\phi}{}_c \, \del_m \ol{b}{}_i \Big) 
\nn \\
\ & \ \ \ \ 
+ \Big\{ 
  \frac{\i \, \ol{a}{}_i \H_{i,-}}{\K_i} \, (\del^m a_i)
+ \frac{\i \, a_i \H_{i,-}}{\K_i} \, (\del^m \ol{a}{}_i)
\Big\}
A_m 
+ \Big\{
\frac{2 \i \, |a_i|^2 a_i \phi_c}{\K_i} \, (\del^m b_i)
- \frac{2 \i \, |a_i|^2 \ol{a}{}_i \ol{\phi}{}_c}{\K_i} \, (\del^m \ol{b}{}_i)
\Big\}
A_m 
\nn \\
\ & \ \ \ \ 
- \frac{\i \, \H_{i,+} \beta_i}{\K_i} \, 
\ve^{mn} A_m \, \del_n (y_i - \ol{y}{}_i)
+ \frac{|a_i|^2 \H_{i,+}}{\K_i} \, 
\big( A_m A^m + 2 |\sigma|^2 \big)
\nn \\
\ & \ \ \ \ 
+ \sqrt{2} \, \H_{i,+} \, D_V 
- \sqrt{2} \, a_i b_i \, D_{c}
- \sqrt{2} \, \ol{a}{}_i \ol{b}{}_i \, \ol{D}{}_{c}
+ \frac{\sqrt{2} \, \beta_i^2}{\K_i} \, |G_i|^2 
+ \frac{\i \sqrt{2} \, \H_{i,+} \beta_i}{\K_i} \, 
\Big\{ 
G_i \sigma - \ol{G}{}_i \ol{\sigma}
\Big\}
\nn \\
\ & \ \ \ \ 
- \frac{(a_i b_i)^2}{\sqrt{2} \, \K_i} \, 
\Big\{
4 F_c M_c 
- 4 G_c N_c 
+ A_{c=} B_{c\+}
\Big\}
- \frac{(\ol{a}{}_i \ol{b}{}_i)^2}{\sqrt{2} \, \K_i} \, 
\Big\{
4 \ol{F}{}_c \ol{M}{}_c 
- 4 \ol{G}{}_c \ol{N}{}_c 
+ \ol{A}{}_{c=} \ol{B}{}_{c\+}
\Big\}
\nn \\
\ & \ \ \ \ 
+ \frac{|a_i|^2 |b_i|^2}{\sqrt{2} \, \K_i} \, 
\Big\{
- 4 |F_c|^2 - 4 |M_c|^2 + 4 |G_c|^2 + 4 |N_c|^2
- A_{c=} \ol{B}{}_{c\+} - \ol{A}{}_{c=} B_{c\+}
\Big\}
\nn \\
\ & \ \ \ \ 
- \frac{|a_i|^2 a_i b_i}{\K_i} \, 
\Big\{
- 2 \sqrt{2} \, \i \, G_c \sigma
+ 2 \sqrt{2} \, \i \, N_c \ol{\sigma}
- A_{c=} (A_0 + A_1)
- B_{c\+} (A_0 - A_1)
\Big\}
\nn \\
\ & \ \ \ \ 
- \frac{|a_i|^2 \ol{a}{}_i \ol{b}{}_i}{\K_i} \, 
\Big\{
2 \sqrt{2} \, \i \, \ol{G}{}_c \sigma
- 2 \sqrt{2} \, \i \, \ol{N}{}_c \ol{\sigma}
- \ol{A}{}_{c=} (A_0 + A_1) 
- \ol{B}{}_{c\+} (A_0 - A_1)
\Big\}
\nn \\
\ & \ \ \ \ 
- \frac{b_i (\K_i + \H_{i,-})}{2 \sqrt{2} \, \K_i} \, 
\Big\{
  4 F_i M_c 
- \i A_{c=} (\del_0 + \del_1) a_i 
- \i B_{c\+} (\del_0 - \del_1) a_i
\Big\}
\nn \\
\ & \ \ \ \ 
- \frac{\ol{b}{}_i (\K_i - \H_{i,-})}{2 \sqrt{2} \, \K_i} \, 
\Big\{
  4 \ol{F}{}_i \ol{M}{}_c 
+ \i \ol{A}{}_{c=} (\del_0 + \del_1) \ol{a}{}_i 
+ \i \ol{B}{}_{c\+} (\del_0 - \del_1) \ol{a}{}_i
\Big\}
\nn \\
\ & \ \ \ \ 
+ \frac{a_i b_i (\K_i + \H_{i,-})}{2 \sqrt{2} \, \ol{a}{}_i \K_i} \, 
\Big\{
  4 \ol{F}{}_i F_c 
+ \i A_{c=} (\del_0 + \del_1) \ol{a}{}_i 
+ \i B_{c\+} (\del_0 - \del_1) \ol{a}{}_i
\Big\}
\nn \\
\ & \ \ \ \ 
+ \frac{\ol{a}{}_i \ol{b}{}_i (\K_i - \H_{i,-})}{2 \sqrt{2} \, a_i \K_i} \, 
\Big\{
  4 F_i \ol{F}{}_c 
- \i \ol{A}{}_{c=} (\del_0 + \del_1) a_i 
- \i \ol{B}{}_{c\+} (\del_0 - \del_1) a_i
\Big\}
\nn \\
\ & \ \ \ \ 
- \frac{a_i (a_i b_i \phi_c + \K_i)}{\sqrt{2} \, \K_i} \, 
\Big\{
   4 \wt{F}_i M_c 
- \i A_{c=} (\del_0 + \del_1) b_i 
- \i B_{c\+} (\del_0 - \del_1) b_i
\Big\}
\nn \\
\ & \ \ \ \ 
- \frac{\ol{a}{}_i (\ol{a}{}_i \ol{b}{}_i \ol{\phi}{}_c + \K_i)}{\sqrt{2} \, \K_i} \, 
\Big\{
  4 \ol{\wt{F}}{}_i \ol{M}{}_c 
+ \i \ol{A}{}_{c=} (\del_0 + \del_1) \ol{b}{}_i 
+ \i \ol{B}{}_{c\+} (\del_0 - \del_1) \ol{b}{}_i
\Big\}
\nn \\
\ & \ \ \ \ 
- \frac{|a_i|^2 \ol{b}{}_i \phi_c}{\sqrt{2} \, \K_i} \, 
\Big\{
  4 \wt{F}_i \ol{F}{}_c 
- \i \ol{A}{}_{c=} (\del_0 + \del_1) b_i 
- \i \ol{B}{}_{c\+} (\del_0 - \del_1) b_i
\Big\}
\nn \\
\ & \ \ \ \ 
- \frac{|a_i|^2 b_i \ol{\phi}{}_c}{\sqrt{2} \, \K_i} \, 
\Big\{
  4 \ol{\wt{F}}{}_i F_c 
+ \i A_{c=} (\del_0 + \del_1) \ol{b}{}_i 
+ \i B_{c\+} (\del_0 - \del_1) \ol{b}{}_i
\Big\}
\nn \\
\ & \ \ \ \ 
+ \frac{a_i b_i \beta_i}{2 \K_i} \, 
\Big\{
4 G_c \ol{G}{}_i + 4 N_c G_i
+ \i A_{c=} (\del_0 + \del_1) (y_i - \ol{y}{}_i)
- \i B_{c\+} (\del_0 - \del_1) (y_i - \ol{y}{}_i)
\Big\}
\nn \\
\ & \ \ \ \ 
+ \frac{\ol{a}{}_i \ol{b}{}_i \beta_i}{2 \K_i} \, 
\Big\{
4 \ol{G}{}_c G_i + 4 \ol{N}{}_c \ol{G}{}_i
+ \i \ol{A}{}_{c=} (\del_0 + \del_1) (y_i - \ol{y}{}_i)
- \i \ol{B}{}_{c\+} (\del_0 - \del_1) (y_i - \ol{y}{}_i)
\Big\}
\nn \\
\ & \ \ \ \ 
+ \text{(fermionic fields)}
\, . \label{LDT-comp}
\end{align}
We note that the scalar field $\phi_c$ again appears in many terms in (\ref{LDT-comp}).
Even though the explicit form (\ref{LDT-comp}) is quite complicated,
it would be useful to study various phases in a generic $\N=(4,4)$ GLSM with an F-term in future works.

\section{$A_1$-type ALE space and its T-duality}
\label{app:EH}

In this appendix we give the explicit form of the $A_1$-type ALE space, 
i.e. the Eguchi-Hanson space \cite{Eguchi:1978gw}, and its T-dualized configuration via the Buscher rule.
We begin with the following form of the Eguchi-Hanson space as a solution of 
supergravity,
\bsubeq \label{EH-config}
\begin{gather}
\d s^2 \ = \ 
\A^{-1} (\d \rho)^2
+ \frac{\rho^2}{4} \Big\{
(\d \vartheta)^2 + \sin^2 \vartheta (\d \varphi)^2
\Big\}
+ \frac{\rho^2 \A}{4} \Big\{ \d \psi + \cos \vartheta (\d \varphi) \Big\}^2
\, , \\
B_{IJ} \ = \ 0
\, , \ls
\e^{2 \Phi} \ = \ 1
\, , 
\end{gather}
\esubeq
where $\A = 1 - \a^4/\rho^4$.
It is interpreted that the Eguchi-Hanson space is equivalent to the two-centered Gibbons-Hawking space \cite{Gibbons:1979zt}.
Actually, the two-centered Gibbons-Hawking metric  
\bsubeq \label{trsf-GH}
\begin{gather}
\begin{align}
\d s^2 \ &= \ 
\Big(\frac{1}{R_+} + \frac{1}{R_-} \Big)^{-1}
\Big\{\d \tau 
+ \Big( \frac{Z_+}{R_+} + \frac{Z_-}{R_-} \Big)
\, \d \tan^{-1} \Big( \frac{Y}{X} \Big) \Big\}^2
\nn \\
\ & \ \ \ \ 
+ \Big( \frac{1}{R_+} + \frac{1}{R_-} \Big)
\big\{ \d X^2 + \d Y^2 + \d Z^2 \big\} 
\, , 
\end{align}
\\
Z_{\pm} \ \equiv \ Z \pm Z_0
\, , \ls
R_{\pm}^2 \ \equiv \ X^2 + Y^2 + Z^2_{\pm}
\, , 
\end{gather}
\esubeq
can be converted to (\ref{EH-config})
by the following transformations \cite{Prasad:1979kg},
\bsubeq
\begin{gather}
X \ = \ Z_0 \sinh \alpha \sin \vartheta \cos \psi 
\, , \ls
Y \ = \ Z_0 \sinh \alpha \sin \vartheta \sin \psi
\, , \ls
Z \ = \ Z_0 \cosh \alpha \cos \vartheta 
\, , \\ 
8 Z_0 \ = \ \a^2 
\, , \ls
8 Z_0 \cosh \alpha \ = \ \rho^2 
\, , \ls
\tau \ = \ 2\varphi
\, . 
\end{gather}
\esubeq
Note that the isometry directions are swapped with
each other by the transformations (\ref{trsf-GH}).  
It is known that the $k$-centered Gibbons-Hawking solution is T-dual to the solution of $k$ parallel NS5-branes along the $\tau$-direction,
where the isometry direction corresponds to the $\varphi$-direction in the Euler coordinate. 
Thus, applying the Buscher rule along the $\varphi$-direction to the metric, the B-field and the dilaton \cite{Buscher:1987} such as 
\bsubeq
\begin{alignat}{3}
G'_{\rho \rho}
\ &= \ 
G_{\rho \rho} 
\, , &\ls
G'_{\vartheta \vartheta}
\ &= \ 
G_{\vartheta \vartheta} 
\, , \\
G'_{\psi \psi}
\ &= \ 
G_{\psi \psi} - \frac{(G_{\varphi \psi})^2}{G_{\varphi \varphi}}
\, , &\ls
G'_{\varphi' \rho}
\ &= \ 
0
\ = \ 
G'_{\varphi' \vartheta}
\, , &\ls
G'_{\varphi' \varphi'}
\ &= \ 
\frac{1}{G_{\varphi \varphi}}
\, , \\
B'_{\varphi' \psi}
\ &= \ 
\frac{G_{\varphi \psi}}{G_{\varphi \varphi}}
\, , &\ls
B'_{\varphi' \rho}
\ &= \ 
0 \ = \ 
B'_{\varphi' \vartheta}
\, , &\ls
\e^{2 \Phi'} 
\ &= \
\frac{\e^{2 \Phi}}{G_{\varphi \varphi}}
\, ,
\end{alignat}
\esubeq
we obtain the T-dualized configuration,
\bsubeq \label{Tdual-EH}
\begin{gather}
\d s^2
\ = \ 
\A^{-1} (\d \rho)^2
+ \frac{\rho^2}{4} (\d \vartheta)^2
+ \frac{\rho^2}{4} \frac{\A \sin^2 \vartheta}{\A \cos^2 \vartheta + \sin^2 \vartheta} 
\, (\d \psi)^2
+ \frac{4}{\rho^2} \frac{1}{\A \cos^2 \vartheta + \sin^2 \vartheta} \, 
(\d \varphi')^2
\, , \\
B'_{\varphi' \psi}
\ = \ 
\frac{\A \cos \vartheta}{\A \cos^2 \vartheta + \sin^2 \vartheta}
\, , \ls
\e^{2 \Phi'}
\ = \ 
\frac{4}{\rho^2} \frac{1}{\A \cos^2 \vartheta + \sin^2 \vartheta} 
\, .
\end{gather}
\esubeq
This configuration represents the two parallel NS5-branes located at $(\rho,\vartheta)=(\a,0)$ and $(\rho,\vartheta)=(\a,\pi)$, or in other words, 
$(X,Y,Z)=(0,0,{\pm}Z_0)$ in the orthogonal coordinates.
One reason why this solution can be interpreted as 
two NS5-branes is that the non-vanishing total flux $\int H = \int \d B$ in the system (\ref{Tdual-EH}) is twice as much as that of a single NS5-brane. 
Since there is a magnetic monopole charge associated with the $\varphi$-direction, the five-brane solution can be obtained by a T-duality transformation along the $\varphi$-direction in (\ref{EH-config}). 
In other words, a T-duality transformation along the $\psi$-direction in (\ref{EH-config}) cannot generate five-brane solutions.

\end{appendix}

}
\end{document}